\documentclass[manuscript,times]{aastex63}
\usepackage{amssymb,amsmath, amsthm, amsfonts, graphics, graphicx}
\usepackage{subfiles}

\begin{document}
\title{Radial Velocity Monitoring of the Young Star Hubble 4: Disentangling Starspot Lifetimes from Orbital Motion\footnote{This paper includes data taken at The McDonald Observatory
of The University of Texas at Austin.}}
\author{Adolfo Carvalho}
\affiliation{Rice University, Department of Physics and Astronomy, Houston, TX 77005}

\author{Christopher M. Johns-Krull}
\affiliation{Rice University, Department of Physics and Astronomy, Houston, TX 77005}

\author{L. Prato}
\affiliation{Lowell Observatory, 1400 Mars Road, Flagstaff, AZ 86001}

\author{Jay Anderson}
\affiliation{STScI, 3700 San Martin Drive, Baltimore, MD 21218}

\begin{abstract}
We studied the weak-lined T Tauri star Hubble 4, a known long-period binary, and its starspot phenomena. We used optical radial velocity (RV) data taken over a span of 14 years (2004-2010, 2017-2019) at the McDonald Observatory 2.7m Harlan J. Smith telescope
and single epoch imaging from the HST/WFC3 instrument.  The observed and apparent RV
variations show contributions, respectively, from the binary motion as well as from a
large spot group on
one of the stars, presumed to be the primary.  
Fitting and removing the orbital signal 
from the RVs, we found the lower bound on the lifetime of 
a previously identified large spot group on the surface of the star to be at least 5.1 years. A $\sim5$ year lower limit is a long, but not unprecedented, duration for a single spot group. The later epoch data indicate significant spot evolution has occurred, placing an upper bound on the spot group lifetime at 12 years. We find that pre-main sequence evolutionary models for the age of Taurus ($\sim2$ Myr), combined with component mass estimates from the literature, permit us to reproduce the HST relative photometry and the binary-induced contribution to the apparent RV variations. The long-lived star spot we
find on Hubble 4 has significant implications for dynamo models in young stars, as it adds evidence for long lifetimes of magnetic field topologies. There are also significant implications for young star exoplanet searches as long-lived coherent RV signals may be spot-induced and not the result of planetary motion.

\end{abstract}


\section{Introduction} \label{sec:intro}

T Tauri stars (TTSs) are young, $\lesssim$ 10 Myr old, $\lesssim$ 2 $M_\odot$, pre-main sequence stars. Some TTSs, such as the classical T Tauri stars (CTTSs), are still surrounded by dusty 
accretion disks, while others, the weak-lined or naked T Tauri stars (WTTSs), no longer 
retain their disks. Studying 
these young stars provides insight into the processes of stellar evolution 
and planet formation and the roles that disk accretion, nearby stellar 
companions, and strong stellar magnetic fields can play in these processes. 
Searching for planets in these young systems provides a snapshot of planet 
formation at its earliest stages. The radial velocity (RV) and transit methods 
commonly used to detect and analyze planets can be confused by spots on the 
surfaces of stars \citep[e.g.,][]{saar_activity-related_1997}. As a result of 
their very strong magnetic fields and large spots 
\citep[e.g.,][]{johns-krull_how_2007}, this can be particularly problematic for 
young stars. A better understanding of the properties, including the 
lifetime, of starspots on young stars can aid in devising strategies to 
mitigate their effects and can give insight into the dynamo processes that 
produce the spots in the first place.

Strong magnetic fields on young stars are of particular interest, as they 
contribute significantly to the star's environment and give clues to the star's 
structure \citep{johns-krull_how_2007, mann_zodiacal_2016}. The appearance of 
large, long-lived spots on the surface of a star are hallmarks of such strong 
magnetic activity. In rapid rotators, spots tend to appear toward the poles, 
resulting in spot formations that can last for many years, as in V410 Tau, 
where a polar spot formation persisted for at least 8 years 
\citep{hatzes_doppler_1995, stelzer_weak-line_2003}, although this may not necessarily 
have been the same spot, but rather may have been similar-looking spot formations dispersing and reappearing 
at the pole \citep{hatzes_doppler_1995}. Similarly, long-lived spots are 
described by \citet{bradshaw_sunspot_2014} on other young, rapidly rotating 
stars. Spots on the surface of slower rotators can remain in lower latitudes 
and contribute to measured Doppler variations of the stars’ spectra more 
strongly than those at higher latitudes, but such spots are thought to rarely 
be coherent for much more than a year \citep{choi_rotation_1996}.

Hubble 4 is a weak-lined T Tauri star in the Taurus-Auriga cloud. It is a 
highly magnetically active K7 star, with a mean surface field strength of 
2.5 kG \citep{johns-krull_testing_2004}, and it was found to have large amplitude 
RV variations attributed to a spot \citep{mahmud_starspot-induced_2011}. In
much of the relevant literature, Hubble 4 is described as a single object; 
however, recent studies have shown it to be a close binary 
system \citep{kraus_mapping_2011, galli_gouldtextquotesingles_2018,rizzuto_dynamical_2020}. Hints 
of this binarity were reported earlier in a study of the radio emission from 
Hubble 4: \citet{skinner_circularly_1993} found the emission was extended in 
such a way that might imply the presence of a companion. 
\citet{kraus_mapping_2011} reported that Hubble 4 is in fact a binary 
system based on Keck Observatory speckle imaging. They find the system 
is composed of two stars, Hubble 4 A and Hubble 4 B, with a measured 
separation of approximately $28.4^{\prime\prime} \pm 0.1^{\prime\prime}$ ($\sim 4.1$ AU). \citet{galli_gouldtextquotesingles_2018} 
used Very Large Baseline Array (VLBA) observations combined with near-infrared 
speckle interferometry of the system to determine dynamical masses for the 
two components, $1.234 \pm 0.023$ $M_\sun$ for the primary 
and $0.730 \pm 0.020$ $M_\sun$ for the secondary. The realization that 
Hubble 4 is a binary system is critical for interpreting RV variations of 
the unresolved system. 

We used RV data and HST/WFC3 photometry of Hubble 4, in conjunction with the system parameters reported by \citet{galli_gouldtextquotesingles_2018}, to study the RV variations produced by orbital 
motion and spot induced signals. Removing the binary motion from the RV curve, we find a coherent starspot signal in the RV data that has remained stable for at least 5.1 years of observations, but less than $\sim$ 12 years. This paper is organized as follows: $\S$ \ref{sec:data} presents the data used in the study, $\S$ \ref{sec:spectroscopy} presents the RV measurements made from the data, $\S$ \ref{sec:analysis} presents the orbital fitting and spot analysis, $\S$ \ref{sec:flare} discusses a fortuitous flare captured in our data, and $\S$ \ref{sec:conclusion} presents our conclusions. Appendix \ref{appendix:app1} provides ancillary data tables and a full analysis of the HST imaging.A physically motivated
RV model of the binary motion is described in Appendix \ref{appendix:2line} which was used for the
orbital RV subtraction in \S \ref{sec:analysis}.

\section{Data} \label{sec:data}
\subsection{Imaging} \label{sec:data_img}
We used the HST images taken through a variety of filters in late 2011 as part of program 12506 (PI-Kraus).  These consist of three exposures through F275W, F336W, F390W, F395N, F438N, F475W, F555W, F625W, F656N, F775W, and F850LP.  The binary was visible with good S/N in each exposure.  We used the ``\_flc" images for this analysis.  These images have been bias-subtracted and flat-fielded.  In addition, a pixel-based correction has been applied to each image to account for charge-transfer inefficiency. A sample image, the third exposure with the F775W filter, is shown in Figure \ref{fig:h4_phot}.

As described in Appendix \ref{sec:imaging}, we analyzed each exposure with an empirical PSF that had been previously extracted from a set of dithered observations of the globular cluster Omega Centauri from program Cal-11452 (PI-Kim-Quijano).  The many point sources in these images made it possible to construct an accurate model of the PSF, including its variation with position.  To analyze each of the Hubble 4 exposures, we extracted a PSF tailored to the particular filter and to the binary's location on the WFC3/UVIS detector.

The three exposures for each filter were taken with the UVIS shutter in different positions. To account for the vibration-induced motion blur from the shutter in position B during shorter exposures \citep{hartig_wfc3_2008}, we apply a Gaussian blur to the PSFs. We use calibration data from \citet{sabbi_uvis_2009} to determine that the best-fitting Gaussian kernel that describes the camera jitter has an RMS of 0.7 pixels and equal elongation in both directions. We applied the blurred PSFs to the exposures taken in Shutter Position B and applied the unblurred PSFs to exposures taken in Shutter Position A. 

\subsection{Spectroscopy}
High resolution optical spectra were obtained using McDonald Observatory's 
2.7 m Harlan J. Smith telescope and the Robert G. Tull cross dispersed 
coud\'e echelle spectrograph \citep{tull_high-resolution_1995}.  Observations 
were taken between the years 2004-2018.  \citet{mahmud_starspot-induced_2011} analyzed
the 26 observations taken between November 2008 and February 2010.  In 
addition to these, we also analyzed 20 observations taken between January 2005
and November 2007 as well as 20 observations taken between November 2017
and January 2019. The dates of the observations are reported in Table \ref{tab:h4_rvs}. Two observations had significant cloud or Lunar contamination so we exclude these from Table \ref{tab:h4_rvs} and from the analysis.  A 1.2$^{\prime\prime}$ slit was 
used to deliver a spectral resolution ($R \equiv \lambda / \Delta\lambda$) 
of $\sim 60,000$. The spectra, cross-dispersed into 54-55 orders, were recorded on a Tektronix $2080 \times 2048$ 
CCD. Observations were made roughly once each night during a 
given observing run (typically 5-10 nights). Before and after each stellar 
spectrum, a Thorium-Argon comparison lamp spectrum was taken to determine the wavelength scale for the observations and detect any instrumental radial velocity shift.  

All spectra were 
reduced with a custom package of IDL echelle reduction routines based 
largely on the data reduction procedures described by 
\citet{valenti_photospheric_1994} and \citet{hinkle_phoenix_2000}.  
The reduction procedure is standard and includes bias subtraction, 
flat fielding by a normalized flat spectrum, scattered light subtraction, 
and optimal extraction of the spectrum.  The blaze function of the echelle 
spectrometer is removed to first order by dividing the observed stellar 
spectra by an extracted spectrum of the flat lamp.  Final continuum 
normalization was accomplished by fitting a 2nd order polynomial to the 
blaze corrected spectra in the regions around the lines of interest for 
this study.  The wavelength solution for each spectrum was determined by 
fitting a two-dimensional polynomial to $n\lambda$ as function of pixel 
and order number, $n$, for approximately 1800 extracted thorium lines 
observed from the internal lamp assembly.

\section{Spectroscopic Analysis} \label{sec:spectroscopy}

The main focus of the work presented in this paper is the analysis of
the high resolution spectroscopy and the actual and apparent RV variations of Hubble 4 that
it reveals.  As described below, our observations imply long lifetimes
for spots on the surface of young stars.  The HST imaging analysis provides
secondary information that complements the main analysis of this paper.
Therefore, to preserve the flow of the main
research presented here, the HST analysis and related results are 
presented in Appendix \ref{appendix:2line}.

\subsection{Radial Velocity Measurements}

\begin{figure}
    \centering
	\includegraphics[angle = 90, origin=c, width= 0.85\linewidth]{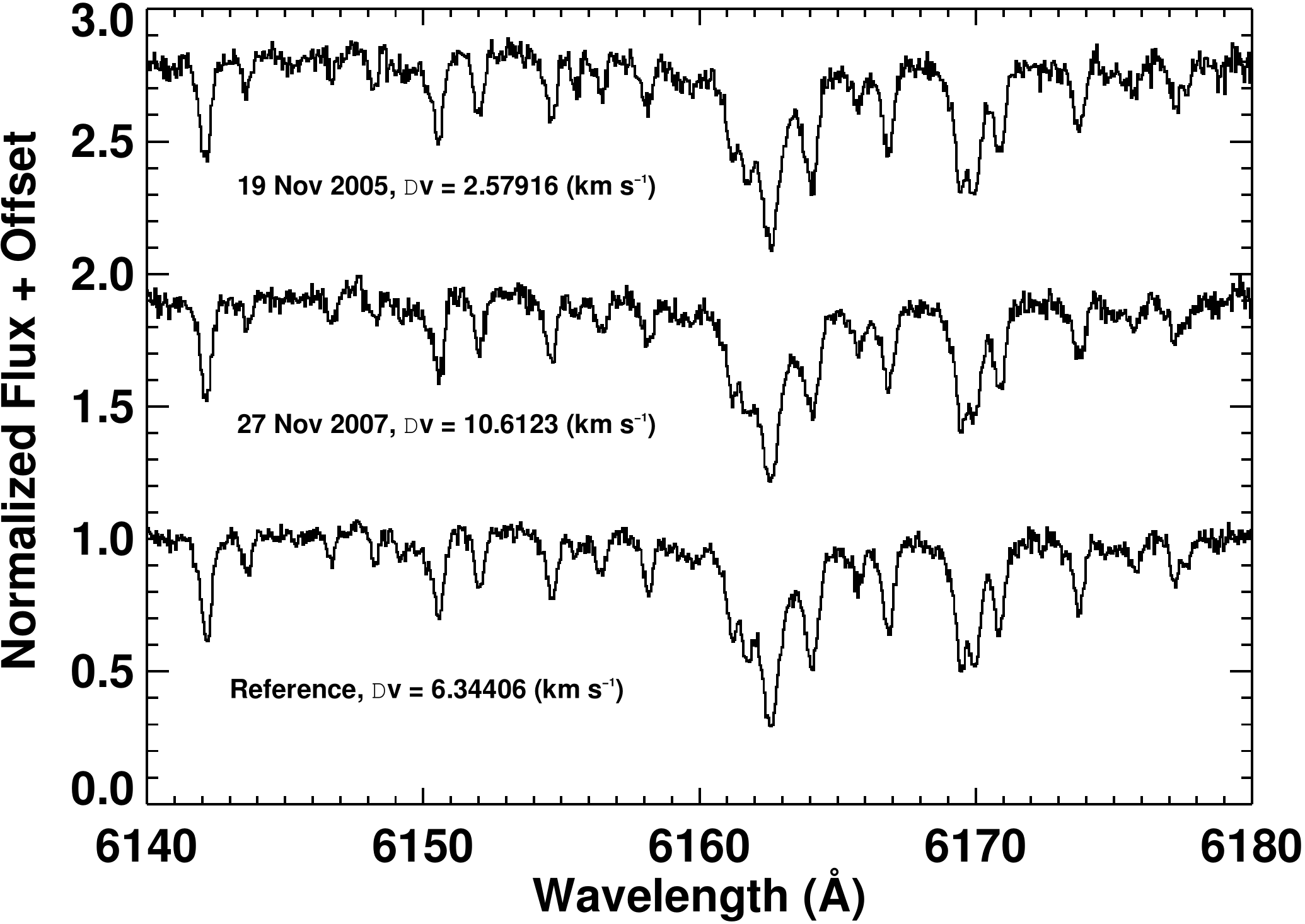}
	\caption{A portion of absolute order number 56 from three selected observations of the Hubble 4 system. The top observation was taken when the binary components were near minimum difference in their velocities. The middle observation was taken near quadrature of the system, when the velocity difference between the components was greatest. The bottom is the reference spectrum, against which radial velocity variations are measured. $\Delta v$ denotes the predicted absolute velocity difference between the two components for each observation.}
	\label{fig:ref_spec_sample}
\end{figure}

We 
determined the RV shifts of the various Hubble 4 spectra relative to one specific spectrum \citep{huerta_starspot-induced_2008, mahmud_starspot-induced_2011}. This reference spectrum was chosen as the observation 
with the highest signal to noise and the fewest noticeable cosmic ray hits 
that survived the reduction process. We chose as our reference the spectrum taken on February 10, 2007 (JD 2454141.6), shown in Figure \ref{fig:ref_spec_sample}. The number of spectral orders used in the analysis is between 9 and 11, depending on the quality of the 
individual spectra. For each observation, every useable order was cross-correlated with the respective order in the reference spectrum \citep{tonry_survey_1979}. The 
Cross-Correlation Function (CCF) between the reference and the observation 
was calculated and 11 pixels around the peak of the CCF were fit with a Gaussian to determine the RV 
shift at the subpixel level. The pixel shift was converted into a velocity 
using the wavelength dispersion and the Doppler formula. To account for 
any instrumental shift, a similar procedure was applied to the Thorium-Argon 
spectra taken before and after each observation and their shifts were 
averaged. This instrumental shift average and the barycentric velocity at the 
midpoint time of observation were then subtracted from the measured RV. 
The resulting value is the RV associated with a given order for a given 
observation. 

The RV values for each order were then averaged, and the uncertainty in the 
final measurement was taken to be the standard deviation of the mean of the 
different RV values from each order, which was added in quadrature to the 
systematic uncertainty in the measurement ($\sim 160$ $\text{ms}^{-1}$; see below). 
To estimate the systematic uncertainty, we applied the same procedure to known
RV standards, $\tau$ Ceti, 107 Psc, and HD 4628. These systems should not 
demonstrate any RV variation above 13 m s$^{-1}$ \citep{fischer_twenty-five_2014},
which is well below the precision we expect to obtain or require.  As a result,
any variation measured in these RV standard stars can be attributed to systematic sources
of uncertainty associated with our measurement technique. We take the 
mean of the standard deviations 
of the RV measurements of the three stars to be the systematic uncertainty. 
The radial velocity measurements for Hubble 4 are shown in Figure 
\ref{fig:og_rvs} and given in Table \ref{tab:h4_rvs} in the Appendix.
The RV values in this new analysis for the observations 
in common with \citet{mahmud_starspot-induced_2011} closely match those 
reported in their paper.  Subtracting one set of RV values from the other
gives a mean difference of 0 km s$^{-1}$ with a standard deviation of the
difference of 200 m s$^{-1}$, which is within the final RV uncertainty 
achieved for our analysis. 

\begin{figure}
    \centering
	\includegraphics[angle = 90, origin=c, width= 0.85\linewidth]{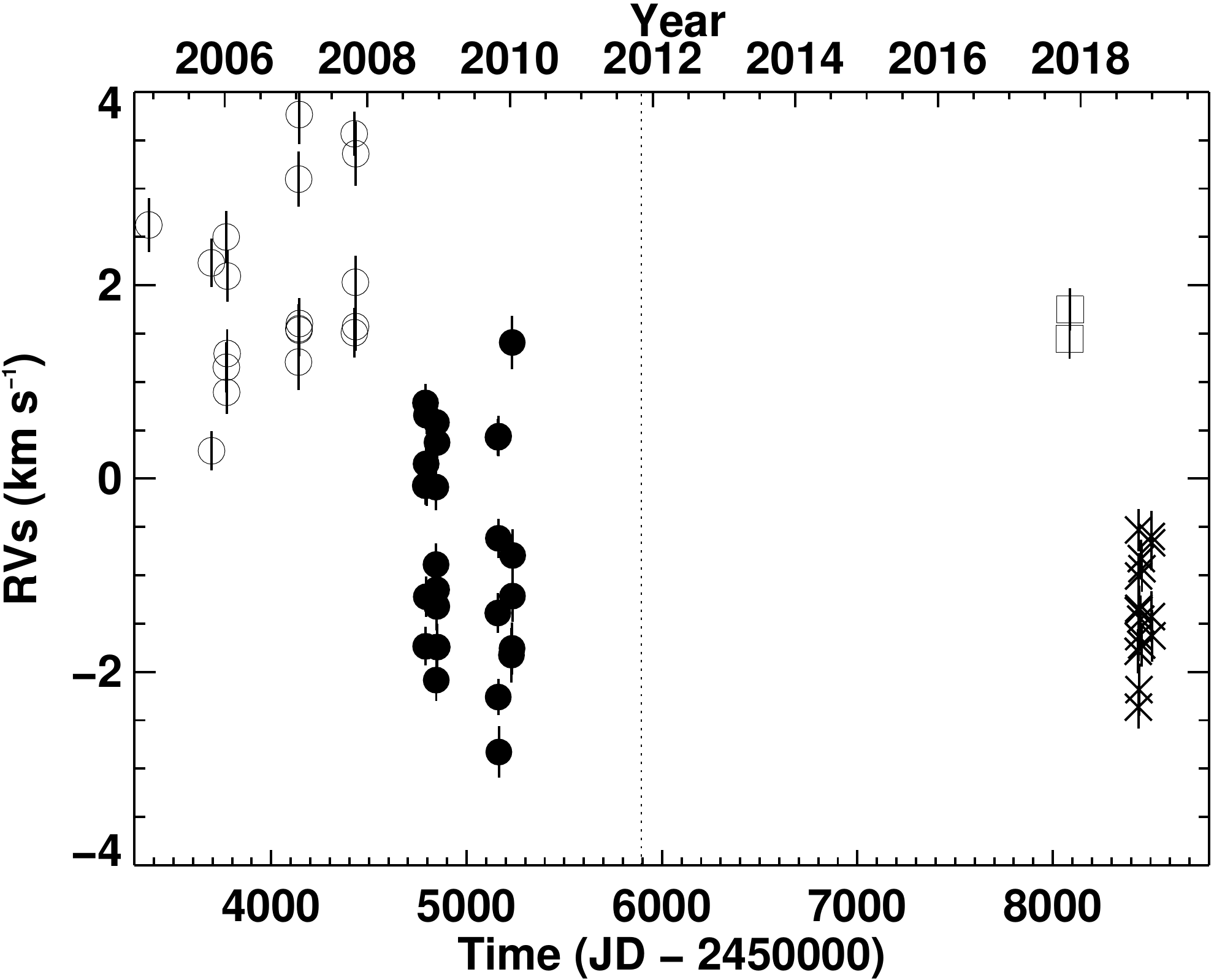}
	\caption{Radial velocity measurements for Hubble 4 obtained with the 2.7m Harlan J. Smith telescope and the 2dcoude optical echelle spectrometer at McDonald Observatory. The 2005-2007 observations are shown with open circles, the 2008-2010 data with filled circles, and the 2018-2019 data with an $\times$. 2017 data which are not included in the analysis are shown with open squares. The dashed line indicates the date of the HST observations.}
	\label{fig:og_rvs}
\end{figure}

\subsection{Line Bisector Measurements}

Starspots can cause spectral line deformations that can 
mimic RV variations in our measurements. Therefore, we are wary 
of any periods resulting from a periodogram analysis that are close to the 
stellar rotation period. Apparent radial velocity variations at these periods might 
be caused by starspots carried across the surface by stellar
rotation.

\begin{figure}
    \centering
	\includegraphics[angle=90, origin=c, width=0.85\linewidth]{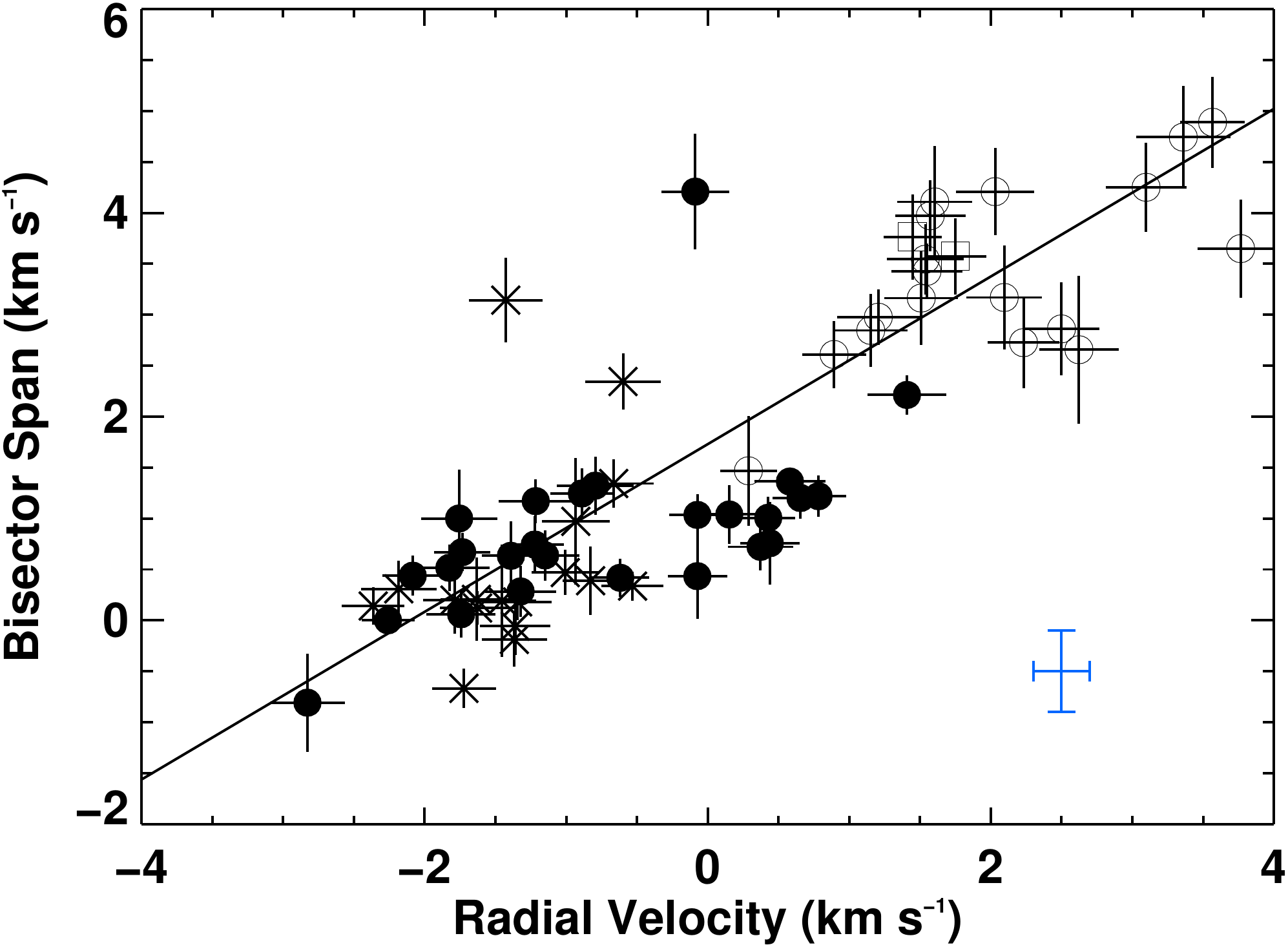}
    \caption{Plot of bisector spans versus radial velocities. The 2005-2007 observations are shown with open circles, the 2008-2010 data with filled circles, and the 2018-2019 data with an $\times$. 2017 data which are not included in the analysis are shown with open squares. The line shows the best fit  linear correlation with slope of 0.74 and false alarm probability of  $2\times10^{-12}$. The blue cross in the lower right corner shows a vertical uncertainty of 400 m s$^{-1}$ and a horizontal uncertainty of 200 m s$^{-1}$ to guide the eye.}
	\label{fig:rvs_vs_bis}
\end{figure}

One way to quantify the deformations of spectral lines and to potentially 
identify signals resulting from star spots is to measure the line bisector spans 
\citep{huerta_starspot-induced_2008}. The bisector is the set of points 
halfway between the two sides of a spectral line profile and the span is 
the inverse mean slope of the bisector. Rather than using individual spectral 
lines, we compute the bisectors of the normalized CCF used to measure the RV. 
The CCF can be considered an average inverse spectral line profile 
\citep{huerta_starspot-induced_2008}. We take two points, one near the top 
of the CCF (mean of the bisector at CCF values 0.8 and 0.9) and one at the 
bottom (mean of the bisector at CCF values 0.15 and 0.25). The slope 
between these two points is then 
calculated. We compute the bisector span of the CCF for each order used to compute the RVs. The bisector spans are then averaged and we take the standard deviation of the mean to be the uncertainty in the span measurement. Strong correlation between the bisector span and 
the RV indicates that the apparent RV variations are likely caused by a spot 
rather than by a massive exoplanetary companion.  Figure \ref{fig:rvs_vs_bis} shows the 
bisector span versus the RV and demonstrates the strong correlation expected for
 spots.  Bisector span measurements 
for Hubble 4 are given in Table \ref{tab:h4_rvs} in the Appendix.

\section{Spot Induced RV Variations and Spot Lifetimes} \label{sec:analysis}

The goal of this paper is to study the lifetime of large starspots on
Hubble 4 by observing the apparent RV signals produced by the spot(s) as the
star rotates, following up on the results of \citet{mahmud_starspot-induced_2011}. 
However, the RV values that we measure in Hubble 4
have two sources: the spot-induced signals we wish to study and also
signals resulting from the binary nature of the Hubble 4 system.  The binary
properties of Hubble 4 are well established \citep{galli_gouldtextquotesingles_2018, rizzuto_dynamical_2020},
therefore, for the purpose of this study, the binary signal needs to be
subtracted out so that the spot-induced signals can be isolated and
studied.  This is done in a two step process.  To accurately fit the 
orbital contribution to the RV variations, the spot induced signals
need to be removed first through an approximate analysis
of the different epochs of our Hubble 4 RV observations.  Once the spot
induced RV signals are approximately removed, we perform an orbital
fit to the residual RV variations.  We then remove the orbital 
contribution from the original measured RVs and perform a more detailed
analysis of the spot induced RV signals.

The HST imaging allowed us to measure the optical flux ratio of
the two members of the binary and that work is presented in detail
in the Appendix \ref{appendix:2line}
\citep[where it is also compared to the recent analysis of][]{rizzuto_dynamical_2020}.
The HST analysis shows that the optical flux ratio $F_2 / F_1$ is $\sim 0.8$, demonstrating that both
stars contribute substantially to the observed optical light.  This
implies that the RVs resulting from the orbit of the binary represent the blending of spectral lines from the two sources. Because this binary has a long period, $\sim 9$ years, and low orbital inclination \citep[$26^\circ$,][Table \ref{tab:h4parms}]{galli_gouldtextquotesingles_2018}, the RVs are low and the individual lines from each star never separate enough in velocity space to
be clearly distinguished. 
The $v$sin$i$ of the components is also large , $\sim$ 14 km s$^{-1}$, compared to the maximum predicted velocity separation of $\sim$ 10 km s$^{-1}$ captured in our observations (Appendix \ref{sec:vsini}).  Appendix \ref{appendix:2line} presents 
detailed modelling of this line blending and the resulting RV signals.

\subsection{Spot Modulation and Orbit Fitting} \label{sec:binary}

\begin{figure}
    \centering
	\includegraphics[angle=90, origin=c, width= 0.85\linewidth]{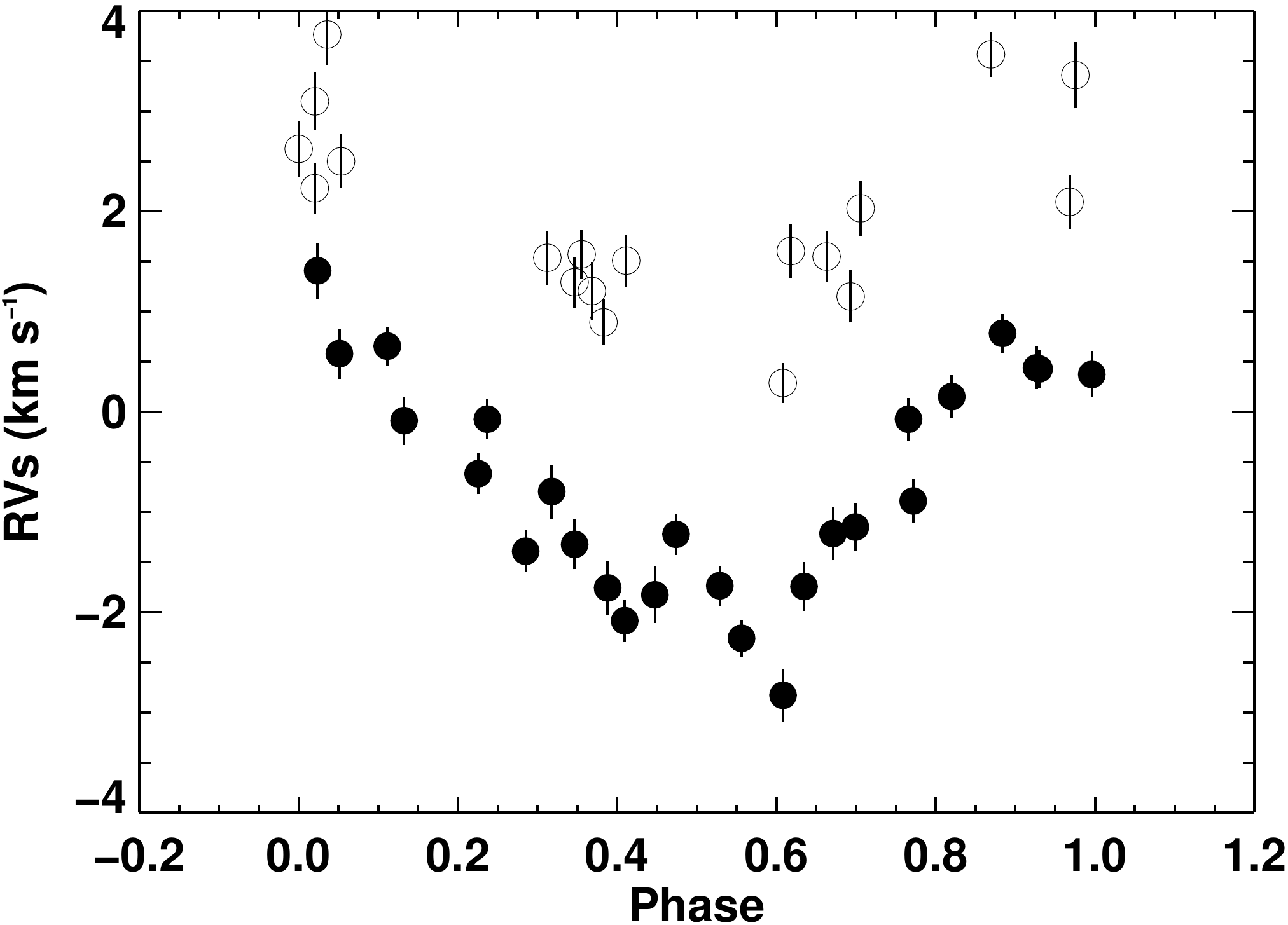}
	\caption{Observations from 2008-2010 and from 2005-2007 phase folded to a period of 1.5459 days. The two sets of measurements have a 2 $\text{km s}^{-1}$ shift between them. Observations from 2005-2007 are shown with open circles and observations from 2008-2010 are shown with filled circles.}
    \label{fig:mahmud_spot}
\end{figure}

\citet{mahmud_starspot-induced_2011} analyzed 26 observations of Hubble 
4 from Nov 17, 2008 to Feb 2, 2010 and found that the RV variations in the 
system were likely the result of the motion of a star spot across the surface of 
the star as it rotates. This conclusion was based on the short (1.5459 days) 
period of the RV variations, the apparent dependence of the amplitude of the 
variations on wavelength (infrared K band observations showed a noticeably 
smaller amplitude), and the strong correlation between the RV variations and 
the bisector spans. Photometric variations of the system's brightness also 
phased to the same period \citep{norton_new_2007}, strengthening the case for 
the presence of a spot. Modeling the spot on the surface of a star, assuming 
a photospheric temperature of 4000K and a spot temperature of 3000K, confirmed that
the presence of a large spot on a star with a rotation period of 1.5459 days 
could reproduce the observed radial velocity semi-amplitude of 
$1.395 \pm 0.094$ $\text{km s}^{-1}$ \citep{mahmud_starspot-induced_2011} in the optical. 
In Figure \ref{fig:mahmud_spot}, the RV variations from 2008-2010 as well 
as from 2005-2007 are shown phased to the same 1.5459 day period.

\begin{figure}%
	\centering
	{\includegraphics[angle = 90, origin=c,width=0.425\linewidth]{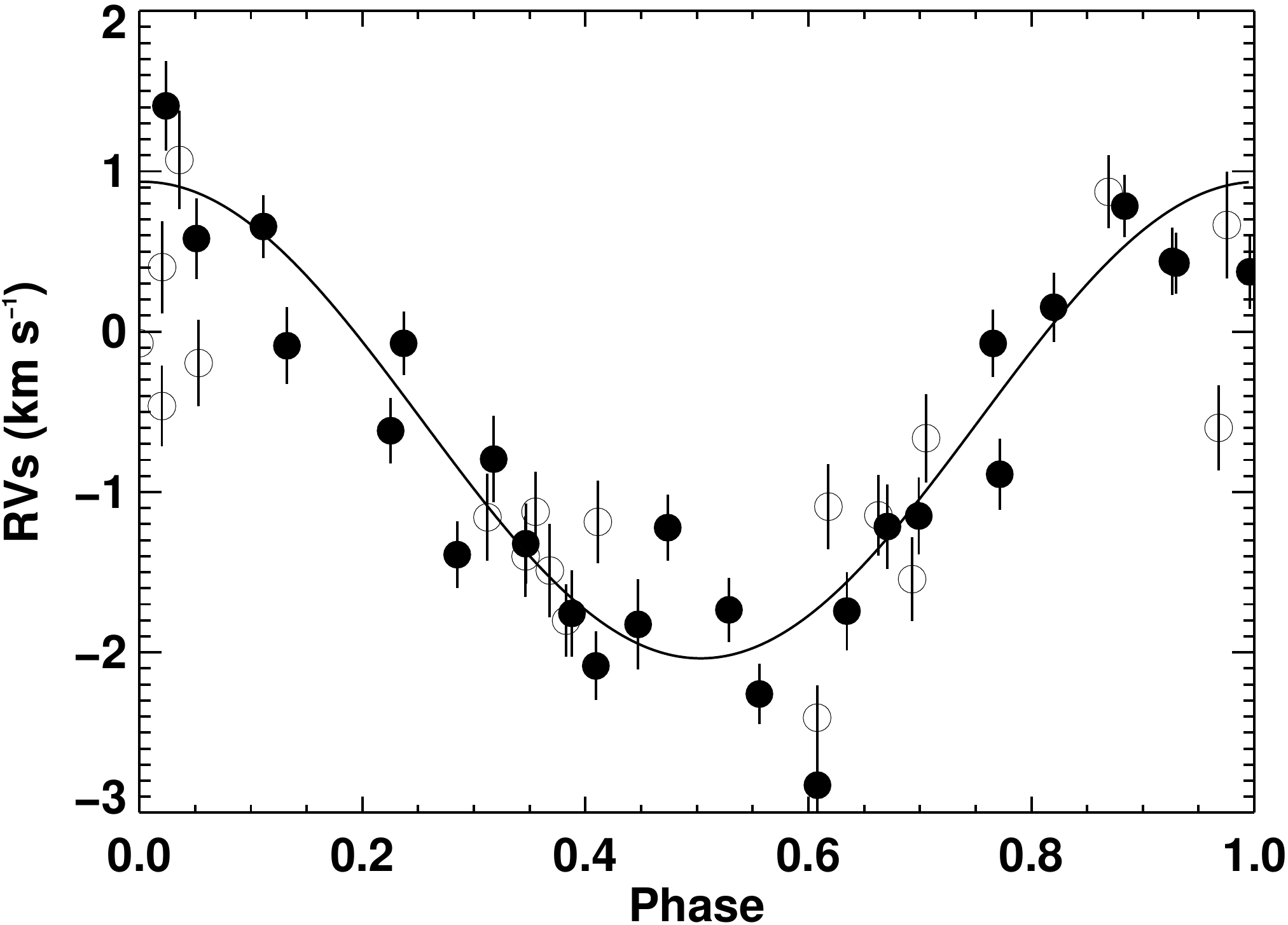} }%
	\qquad
	{\includegraphics[angle = 90, origin=c,width=0.425\linewidth]{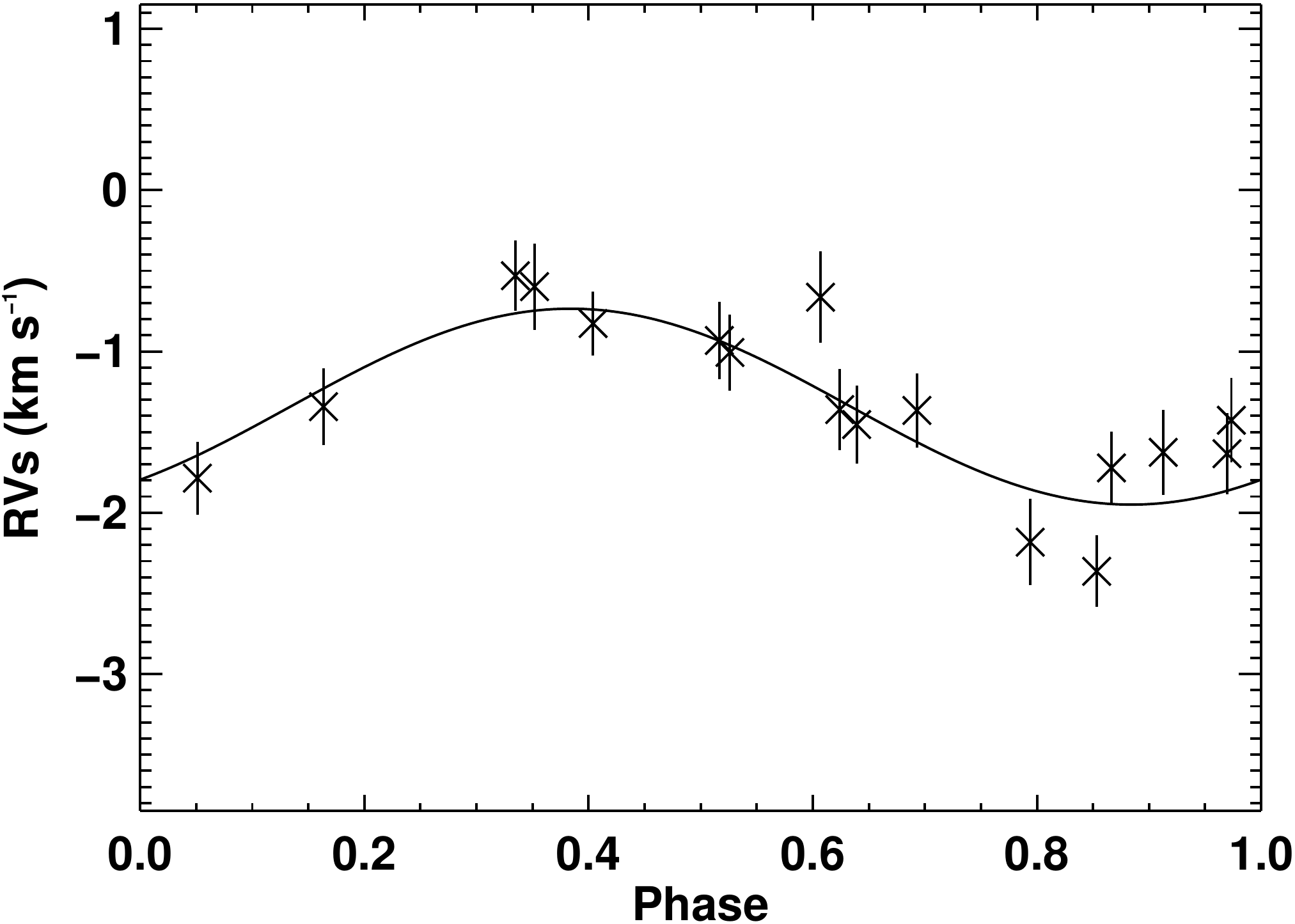} }%
	\caption{Two coherent spot signatures are present in the Hubble 4 RV data phased to a 
	period of 1.5459 d. Solid lines show the sinusoidal fits to the spot-induced RV modulation, used to remove the spot signals from the full RV curve before modeling the binary. \textbf{Left:} Mean subtracted 2008-2010 data and mean subtracted 2005-2007 data. The 2005-2007 observations are shown with open circles, and the 2008-2010 data are shown with filled circles. After subtraction of the means (difference of 2 $\text{kms}^{-1}$) the two sets are indistinguishable. \textbf{Right:} Data from 2018-2019 demonstrating a spot signal with the same period as the spot present in the 2005-2007 and 2008-2010 data, but shifted in phase and with smaller amplitude. Data from 2017 are not coherent with either spot signal and are omitted from these plots.}%
	\label{fig:example_shift}%
\end{figure}

Figure \ref{fig:mahmud_spot} shows that the 2005-2007 RV values are shifted roughly +2 km s$^{-1}$ 
relative to the later dates presented by \citet{mahmud_starspot-induced_2011}.  There is a clear 
separation apparent for the bisector spans as well
(Figure \ref{fig:rvs_vs_bis}). Subtracting this apparent RV shift, the two sets 
of observations show coherent variations, as illustrated in the left panel of 
Figure \ref{fig:example_shift}. This implies that the spot responsible for the RV variations may have 
been stable for over 5 years, a relatively long lifetime for a star spot although not unprecedented \citep{stelzer_weak-line_2003}. 
The most recent data, from 2017-2019, also phase well to the 1.5459 day period but 
are not consistent in phase and amplitude with the older data (right panel of Figure 
\ref{fig:example_shift}).  At first glance, this suggests 
significant spot evolution between the two epochs.

\begin{deluxetable}{ccc}
	\tablecaption{Hubble 4 Orbital Elements  \citep{galli_gouldtextquotesingles_2018}\label{tab:h4parms}}	

	\tablewidth{0pt}
	\tablehead{
		\colhead{Parameter} & \colhead{Value} & \colhead{Uncertainty}
	}
	\startdata{}
	$a$ (mas) & $40.0$ & $0.3$ \\
	$a_1$ (mas) & $16.0$ & $0.3$ \\
	$a_2$ (mas) & $27.0$ & $0.3$ \\
	$M_{total}$ ($M_{\sun}$) & $1.964$ & $0.033$ \\
	$m_1$ ($M_{\sun}$) & $1.234$ & $0.023$ \\
	$m_2$ ($M_{\sun}$) & $0.730$ & $0.020$ \\ 
	$P$ (years) & $9.329$ & $0.017$ \\ 
	$T_P$ (JD) & $2454712$ & $3$ \\
	$e$ & $0.682$ & $0.003$ \\ 
	$i$ (deg) & $153.8$ & $1.2$ \\ 
	$\Omega$ (deg) & $66.1$ & $2.3$ \\ 
	$\omega$ (deg) & $70.0$ & $2.2$  \\
	\enddata
\end{deluxetable}

\begin{figure}
    \centering
	\includegraphics[angle=90, origin=c, width= 0.85\linewidth]{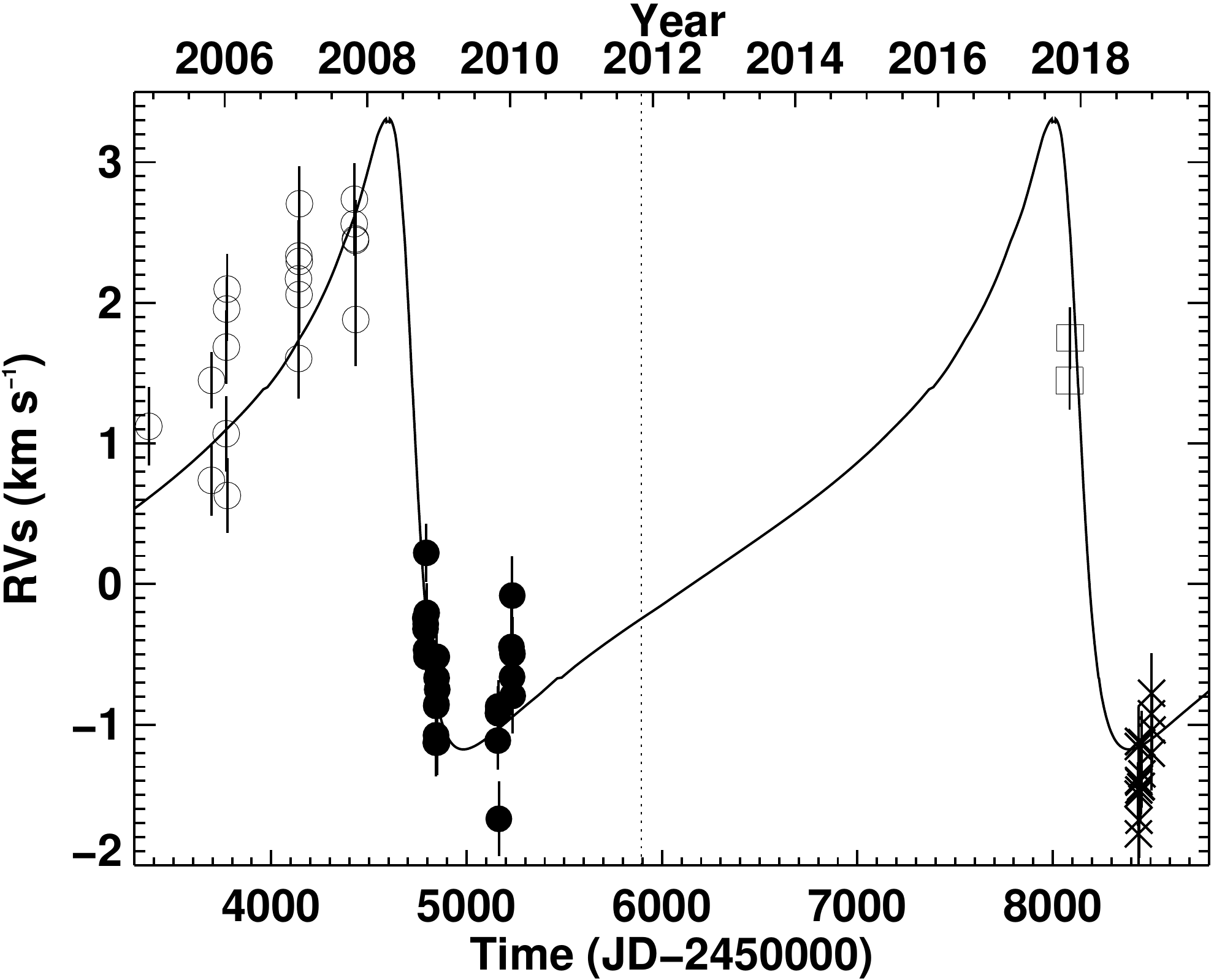}
	\caption{Filled circles show the Hubble 4 RV measurements after subtraction of the sinusoidal fits resulting from spot-induced apparent RV variations shown in Figure \ref{fig:example_shift}.  The solid line shows the blended, double line fit to the orbit using the parameters
	from \citet{galli_gouldtextquotesingles_2018} described in Appendix \ref{appendix:2line}.  The dotted line
	indicates the date of the HST observations. Though the 2017 data are shown in the figure for reference, they are omitted from the blended line fitting procedure.}
	\label{fig:rv_curve_model}
\end{figure}

The simplest explanation for the RV offset seen in Figure \ref{fig:mahmud_spot}
is an instrumental shift that is somehow unaccounted for by the Thorium-Argon 
spectra. However, a purely instrumental explanation is unlikely because it 
appears in the bisector spans as well (Figure \ref{fig:rvs_vs_bis}). The bisector spans are 
computed differentially and do not depend on the absolute wavelength 
calibration. Thus, we conclude the offsets in the RVs and bisector spans are 
real and potentially reflect the motion resulting from the long period binary.
\citet{galli_gouldtextquotesingles_2018} used their own VLBA observations combined 
with Keck/NIRC2 data \citep[later published by][]{rizzuto_dynamical_2020} to determine 
the orbital motion of the binary and compute the orbital parameters of 
the system, summarized in Table \ref{tab:h4parms}.  We use these 
parameters to fit the orbital contribution to the RV variations of Hubble 4.  In order
to isolate the orbital contribution from the spot-induced RV variation, we fit the
two panels of data in Figure \ref{fig:example_shift} with the simple sine waves shown in the plots and subtract these
fits from the RV measurements in Figure \ref{fig:og_rvs}. The two measurements from 2017 do not phase well with either spot signal, so they are omitted from the sine wave subtraction and spot fits shown in Figure \ref{fig:example_shift} and Figure \ref{fig:spot_model_plots}. These initial spot-subtracted RV
measurements are shown in Figure \ref{fig:rv_curve_model}. We then fit a double-lined spectroscopic binary model to these RV measurements (Appendix \ref{appendix:2line}), holding the parameters fixed at the values given in Table \ref{tab:h4parms}. The only free parameters of the fit are the flux ratio of the components and their spectra line depth ratios. Due to the difficulty in phasing the 2017 data with the other RVs, we omit these points from the binary model fit. The observed RV variations
actually result from the blended spectra of the two stars. Using the orbital parameters from Table \ref{tab:h4parms} yields a peak-to-peak RV amplitude of $\sim 8$ km s$^{-1}$ for the primary and $\sim 13.5$ km s$^{-1}$ for the secondary.  The maximum predicted RV separation between the two components at the times we observed Hubble 4 is $\sim 11$ km s$^{-1}$, which is less than the $v$sin$i$ of $\sim14$ km s$^{-1}$ (see Appendix \ref{sec:vsini}) for the components.  These predicted RV amplitudes are significantly larger than that seen in Figure \ref{fig:rv_curve_model}, again, because the orbital signal seen in this figure results from the blending of the lines from the two components, and what we measure is the shift of the apparent velocity of these blended lines.
 We performed a double-lined analysis, described in Appendix \ref{appendix:2line}, and report those results there;  this is
shown as the solid line in Figure \ref{fig:rv_curve_model}. We used a fit to our blended line model to subtract the orbital
motion from the RV variations shown in Figure \ref{fig:og_rvs} and used these orbital RV-subtracted data
to further study the spot-induced RV signals of Hubble 4 (\S \ref{sec:spot rvs}).

\subsection{Spot Lifetime and Model Fits} \label{sec:spot rvs}

Subtracting our blended-line binary model (solid line in Figure \ref{fig:rv_curve_model}) from
the observed RVs precisely removes the offset between RV measurements 
shown in 
Figure \ref{fig:mahmud_spot} (Figure \ref{fig:spot_model_plots}). This correction for
the orbital velocity also appears to preserve the phase coherence of the spot-induced RV
modulation observed in 2008-2010 
compared to the 2005-2007 observations.  This suggests that the
spot or spot complex responsible for these RV modulations maintained its
basic size and location for 5 years (2005 - 2010) or more.

Data from 2018-2019 appear to phase well to the 
same rotation period as the earlier data, but there appears to be
a significant phase shift  (and amplitude change) compared with the 2005-2010
observations (Figure \ref{fig:example_shift}). To test whether
this apparent phase shift is real, or whether a single period could be found
which phased all the data together, we performed a dense search of nearby periods 
around the nominal rotation period (1.5459 d) in an effort to see if a
single period could bring all the RV data into phase. The RV data appear very
sinusoidal in nature (Figure \ref{fig:example_shift}), so we also evaluated different 
possibilities through a reduced $\chi^2$ analysis of sine wave fits.  As mentioned in Section \ref{sec:binary},
the two data points from November 2017 do not appear to phase well with 
any of the other data, so we evaluate the fits without these
two points.  

The RV measurements from 2005 - 2010 provide more sensitivity to the rotation
period than the 2018 - 2019 data because the earlier epochs have significantly
more measurements (45 compared to 17) spread out over a significantly longer time
interval (1862 d compared to 71 d).  Therefore, we first determined how precisely
we can measure the period using only the 2005 - 2010 data and then see how the
later data phases to the range of allowed periods.  To estimate the period in
the 2005 - 2010 data, we performed 10,000 Monte Carlo simulations of the RV 
measurements, determining the period for each iteration.  We performed two sets
of Monte Carlo simulations: the first determined the period using the Lomb
Scargle periodogram \citep[][]{horne1986} and the second determined the
period by minimizing $\chi^2$ for a sine wave fit to the data points.  For
each iteration of the Monte Carlo runs, a new data set was created using the
measured RV values and adding normally distributed random noise where the
$\sigma$ value from the normal distribution was set equal to the measurement
uncertainty for the measured RV points.  The best period estimate is taken as
the mean of the 10,000 Monte Carlo runs and the period uncertainty is taken as
the standard deviation of periods from the Monte Carlo runs.  The Lomb Scargle
method returned a period of $1.545909 \pm 0.000042$ d and the minimizing $\chi^2$
method returned a period of $1.545900 \pm 0.000043$ d.  We average these 
determinations to get a final period estimate of $P_0 = 1.5459045 \pm 0.0000430$ d.
The nominal period (1.5459 d) cited above (and used to compute the phases in
Figures 5 and 8) is only 0.1$\sigma$ different than our new period, $P_0$.  The
2018 - 2019 data is measured approximately 3000 d after the end of the 2005 - 2010 
data.    Using our new period and uncertainty, this 3000 d corresponds to 1940
rotation periods and results in a potential phase error of 0.05, indicating that
we can meaningfully test whether the two epochs maintain phase coherence.

\begin{deluxetable}{cccccc}
	\tablecaption{Sine Wave Fits to Spot Properties\label{tab:sin_spot}}  
	\tablewidth{0pt}
	\tablehead{
		\colhead{ } & \colhead{A$_1$} & \colhead{A$_2$} & \colhead{$|\Delta \phi|$} &
		\colhead{ } & \colhead{ } \\
		\colhead{Period} & \colhead{(km s$^{-1}$)} & \colhead{(km s$^{-1}$)} & \colhead{(0.0 - 0.5)} &
		\colhead{$\chi_r^2$} & \colhead{$P_{\chi^2}$}
	}
	\startdata{}
	$P_0$ & $1.146 \pm 0.062$ & $0.496 \pm 0.103$ & $0.377 \pm 0.040$ & 1.618 & 0.00253 \\
	$P_0 + \sigma_P$ & $1.146 \pm 0.063$ & $0.496 \pm 0.104$ & $0.311 \pm 0.040$ & 1.642 & 0.00189 \\
	$P_0 - \sigma_P$ & $1.145 \pm 0.063$ & $0.496 \pm 0.103$ & $0.443 \pm 0.040$ & 1.634 & 0.00210 \\
	$P_0 + 2\sigma_P$ & $1.146 \pm 0.064$ & $0.496 \pm 0.106$ & $0.244 \pm 0.040$ & 1.709 & 0.00083 \\
	$P_0 - 2\sigma_P$ & $1.145 \pm 0.064$ & $0.496 \pm 0.105$ & $0.490 \pm 0.041$ & 1.690 & 0.00104 \\
	$P_0 + 3\sigma_P$ & $1.145 \pm 0.067$ & $0.496 \pm 0.108$ & $0.177 \pm 0.052$ & 1.819 & 0.00020 \\
	$P_0 - 3\sigma_P$ & $1.143 \pm 0.066$ & $0.496 \pm 0.108$ & $0.423 \pm 0.042$ & 1.790 & 0.00030 \\
	\enddata
\end{deluxetable}

It is clear in both Figure 5 and 8 that the 2018 - 2019 RV data points appear
shifted in phase and have a substantially lower amplitude than the 2005 - 2010
data.  In order to quantify the changes in phase and amplitude, we fit sine 
wave functions to the two sets of RV points, holding the period fixed to the
same value for both fits.  In order to evaluate the quality of the fits, we
compute the value of reduced $\chi^2$, $\chi_r^2$, as well as the probability, 
$P_{\chi^2}$, 
of obtaining the value of $\chi_r^2$ this high or higher from a good fit to the
data.  When calculating these values of $\chi_r^2$ we use the measured RV 
uncertainty values.  However, as can be seen from the $\chi^2$ values reported in
Table \ref{tab:sin_spot}, the values of $\chi_r^2$ even for the best fit are a
little high, indicating that either the measurement uncertainties are slightly
underestimated (by a factor of $\sim 1.27$) or that the spot-induced RV signal
is not perfectly sinusoidal.  Therefore, to quantify the uncertainty in the
amplitude and phase of each sine wave, we multiply the measured uncertainties
by a constant value equal to $\chi_r$ - the square root of the reduce $\chi^2$
value - so that the new value of $\chi_r^2 = 1.0$.  In Table \ref{tab:sin_spot} we
report the amplitude of the two sine waves, $A_1$ (for 2005 - 2010) and $A_2$ (for
2018- 2019), as well as the absolute value of the shift in phase, $|\Delta \phi|$ 
between the two sine waves.  Table \ref{tab:sin_spot} gives the results of these
fits for periods of $P_0$ and $P_0 \pm a \sigma_P$ for values of $a = 1, 2, 3$.
In all cases the amplitude of the two sine waves are significantly ($\sim 5.4\sigma$)
different and the phase difference is always greater than $3\sigma$ (for $P_0$ the
phases are different at the $9.4\sigma$ level).  The phases of the two sine waves
are closest for a period of $P_0 + 3 \sigma_P$, but they are still different at the
$3.4\sigma$ level.  Furthermore, the value of $P_\chi^2$ indicates this fit is a
factor of more than 10 less probable than the nominal $P_0$ fit.  As a result of
all these tests, we conclude that the spot structure evolved significantly
between the earlier and later epochs of RV data.
While we cannot be sure the spot structure did not change appreciably
during 2005-2010, the RV data we have collected are consistent with a very
stable spot or spot complex during this time.  Thus, we suggest that the
lifetime of this spot or spot complex is at least as long as the time
between the first and last observation in this period, corresponding to
a minimum spot lifetime of $\sim 5.1$ years. Given the phase evolution of the spot
in the later data, we can identify also an upper limit on the lifetime of this 
particular spot formation at $\sim 12$ years. This
is consistent with the longest spot lifetimes observed on other young stars \citep{hatzes_doppler_1995, stelzer_weak-line_2003-1, bradshaw_sunspot_2014, robertson_persistent_2020}.

\begin{figure}%
	\centering
	{\includegraphics[angle = 90, origin=c,width=0.85\linewidth]{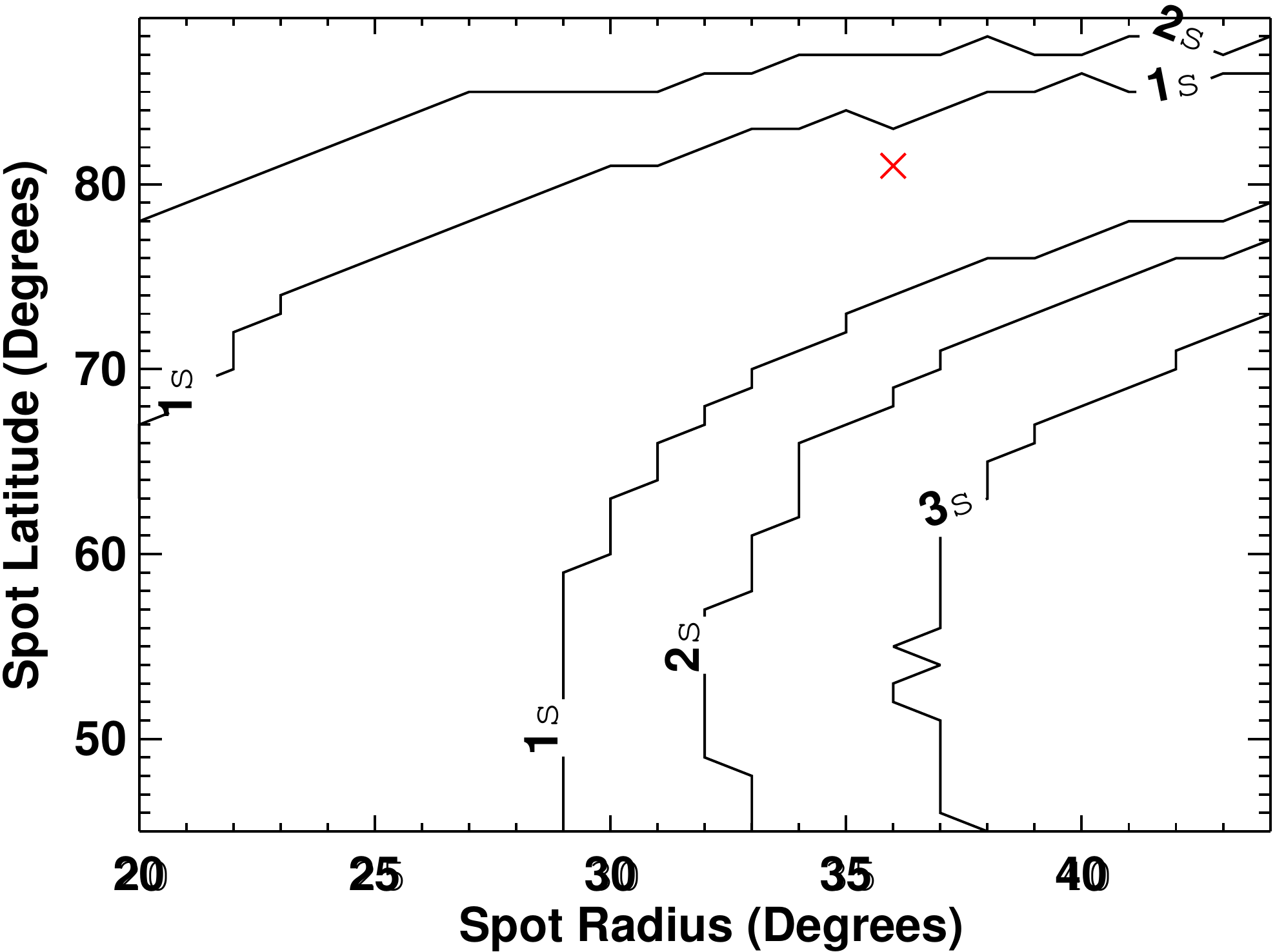} }
    \caption{Contour plot showing $1\sigma$, $2\sigma$, and $3\sigma$ contours for the chi-squared surface for spot model fits to the 2005-2010 RV data. The radii and latitudes are sampled in increments of 1$^\circ$ The red "x" marks the best fit parameters.}%
	\label{fig:spot_contour}%
\end{figure}

\begin{figure}%
	\centering
	{\includegraphics[angle = 90, origin=c,width=0.425\linewidth]{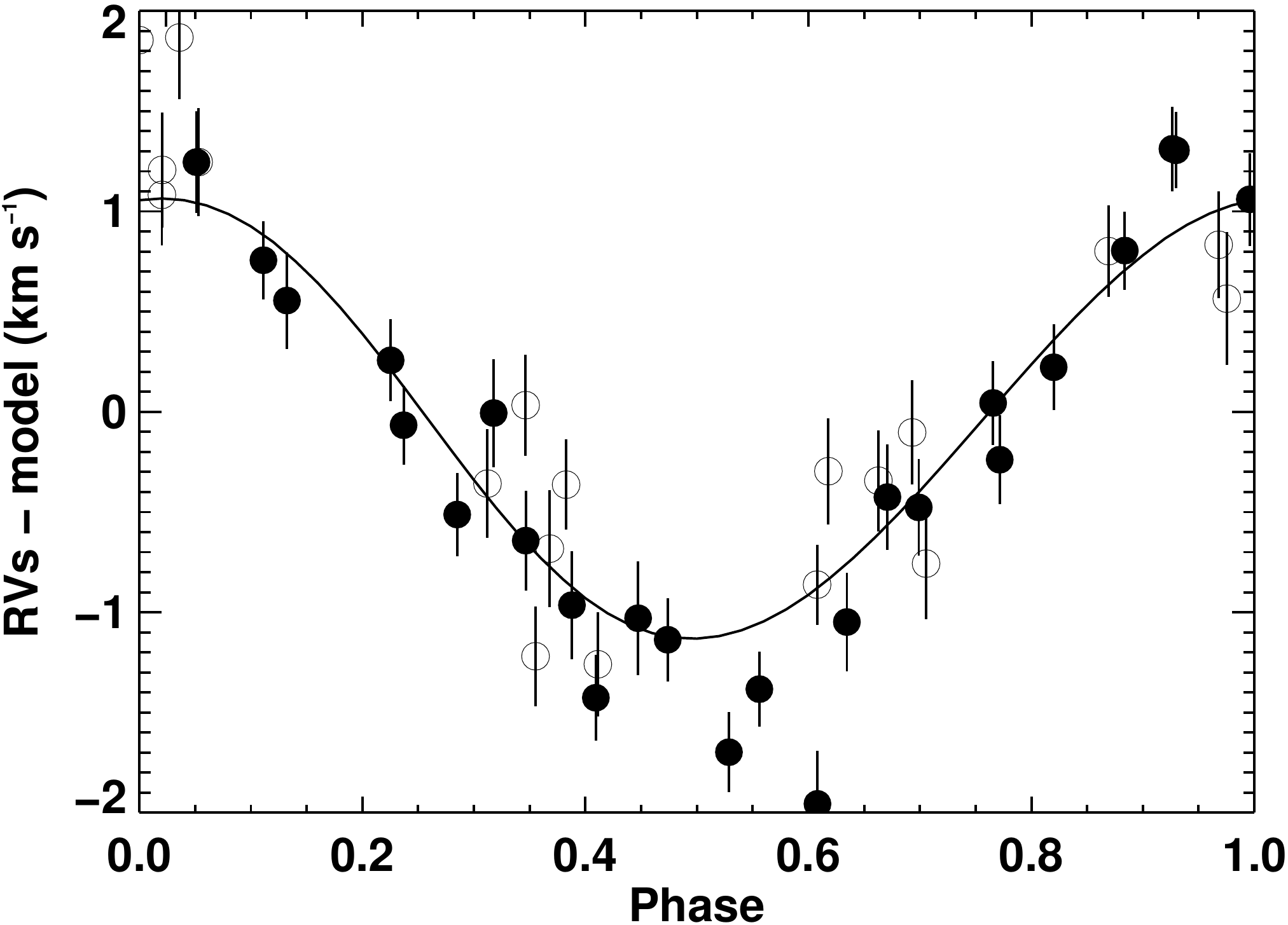} }
	\qquad
	{\includegraphics[angle = 90, origin=c,width=0.425\linewidth]{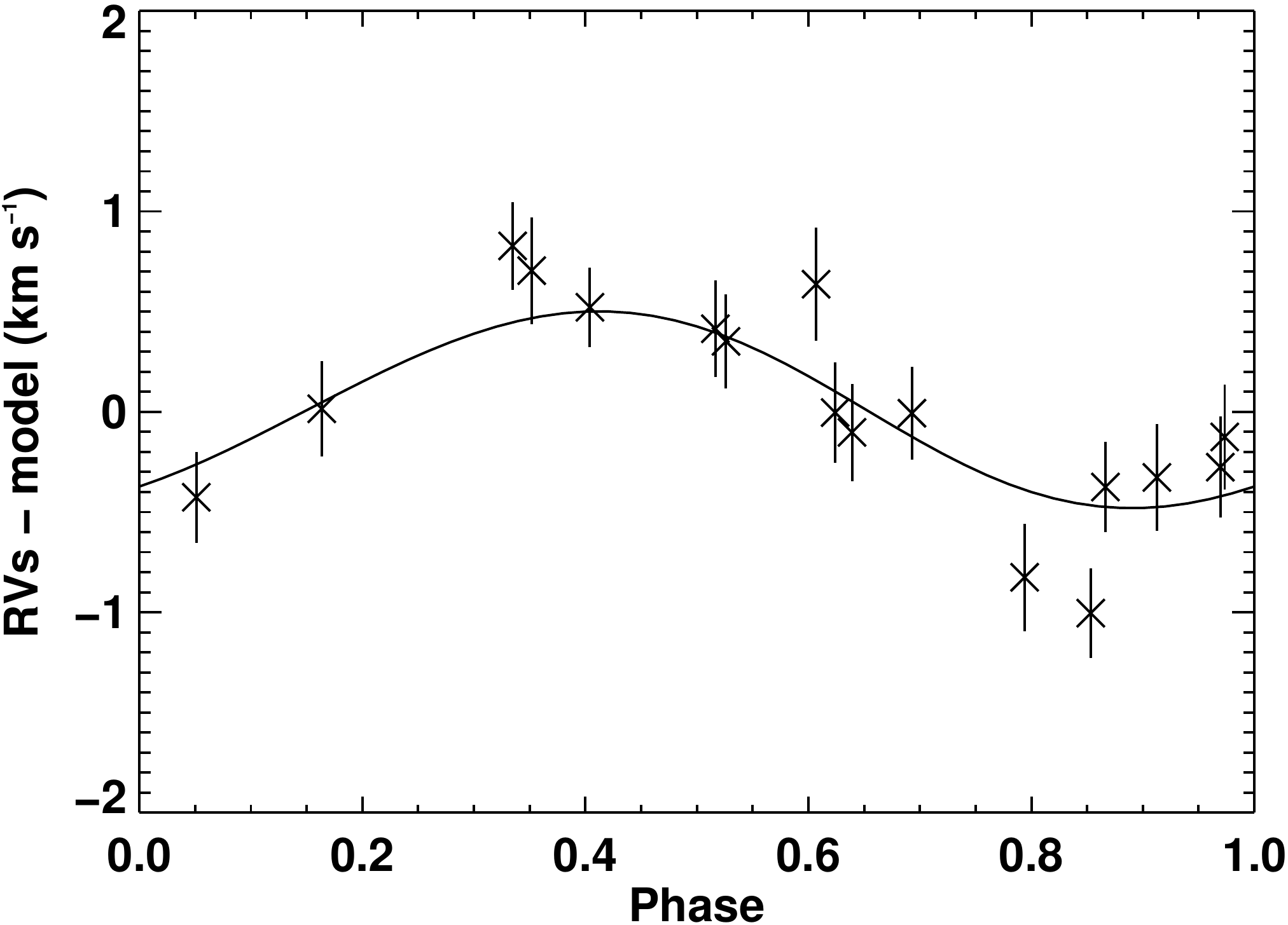} }
	\caption{Orbital motion-subtracted RV curves for the observed spot signals. \textbf{Left:} The 2005-2007 data (shown with open circles) and the 2008-2010 data (shown with closed circles) , which are almost indistinguishable after subtracting the binary RV contribution. The solid line shows a circular spot model assuming a radius of 36$^\circ$ on the star and a latitude of 81$^\circ$ overlaid with the RV data phased to P = 1.5459 days for the 2005-2010 spot. \textbf{Right:} Spot model assuming a radius of 24$^\circ$ on the star and a latitude of 81$^\circ$, in addition to a longitudinal shift of 160$^\circ$ (solid line) with respect to the 2005-2010 data overlaid on the 2018-2019 RV data phased to P = 1.5459 days. }%
	\label{fig:spot_model_plots}%
\end{figure}

Since it appears Hubble 4 has very long-lived spots on its surface, we
investigated further the nature of these spots.
We used the RV measurements to fit a spot model to the RV curve corrected 
for orbital motion. Applying a disk integration procedure similar to that described in 
\citet{huerta_starspot-induced_2008}, we determined the latitude, phase, and
size of a single circular star spot that produces the best fit for the RV
signal from 2005 to 2010, and we determined separate parameters 
for the spot from 2018 to 2019. To identify the best fit, we densely sampled a range of latitudes, phases, and sizes of a single circular spot and computed the chi-squared surface to identify the parameters at which chi-squared was minimum. For the inclination of the star, we adopted a value of $11.8^\circ$, computed assuming stellar radius of 2.1 $R_\odot$ from the \citet{baraffe_new_2015} isochrone for a 2 Myr old $1.2 M_\odot$ star, a 1.5459 day rotation period and a $v$sin$i$ of 14 km
s$^{-1}$ (see Appendix \ref{sec:vsini}). The stellar inclination we assume differs from 
the inclination of the orbital plane by $\sim 14^\circ$.  While it might be expected that the 
orbital and stellar inclinations should be the same, a number of examples of binary
young stars have been found where circumstellar disk inclinations are misaligned 
relative to the binary orbital inclination \citep[e.g.][]{Kurt2018}, and circumbinary disks have
been discovered which are misaligned relative to the binary orbit \citep[e.g.][]{Cze2019}.  As
a result, it appears that the minor misalignment we suggest for Hubble 4 is far from unique.
While the phase is well-determined, unfortunately the relationship between the 
latitude and size of the spot is highly degenerate as a result of the very low $11.8^\circ$
stellar inclination, thus there are multiple combinations of spot radii 
and latitudes which convincingly mimic the spot signal. The contour plot in Figure \ref{fig:spot_contour} demonstrates the degeneracy between the choices of radius and latitude values to fit the 2005-2010 RV data. The formal best fit 
parameters indicate a spot radius of 36$^\circ$ at a latitude of 81$^\circ$ for 2005-2010 (left panel of Figure \ref{fig:spot_model_plots}). At 
that same latitude, the best fit radius for the 2018-2019 spot is 24$^\circ$
with a shift of $\sim 160^\circ$ in longitude (right panel of Figure 
\ref{fig:spot_model_plots}). The spot coverage of the surface of the star from 
these spots agrees well with that found from the analysis of the HST 
imaging. 
We are confident that the more recent signal represents a smaller spot
at any latitude and a longitude which differs by $\sim 160^\circ$. Completely different spot groups dominated on Hubble 4 during these two epochs.

\begin{figure}
    \centering
	\includegraphics[ angle = 90, width= 0.85\linewidth]{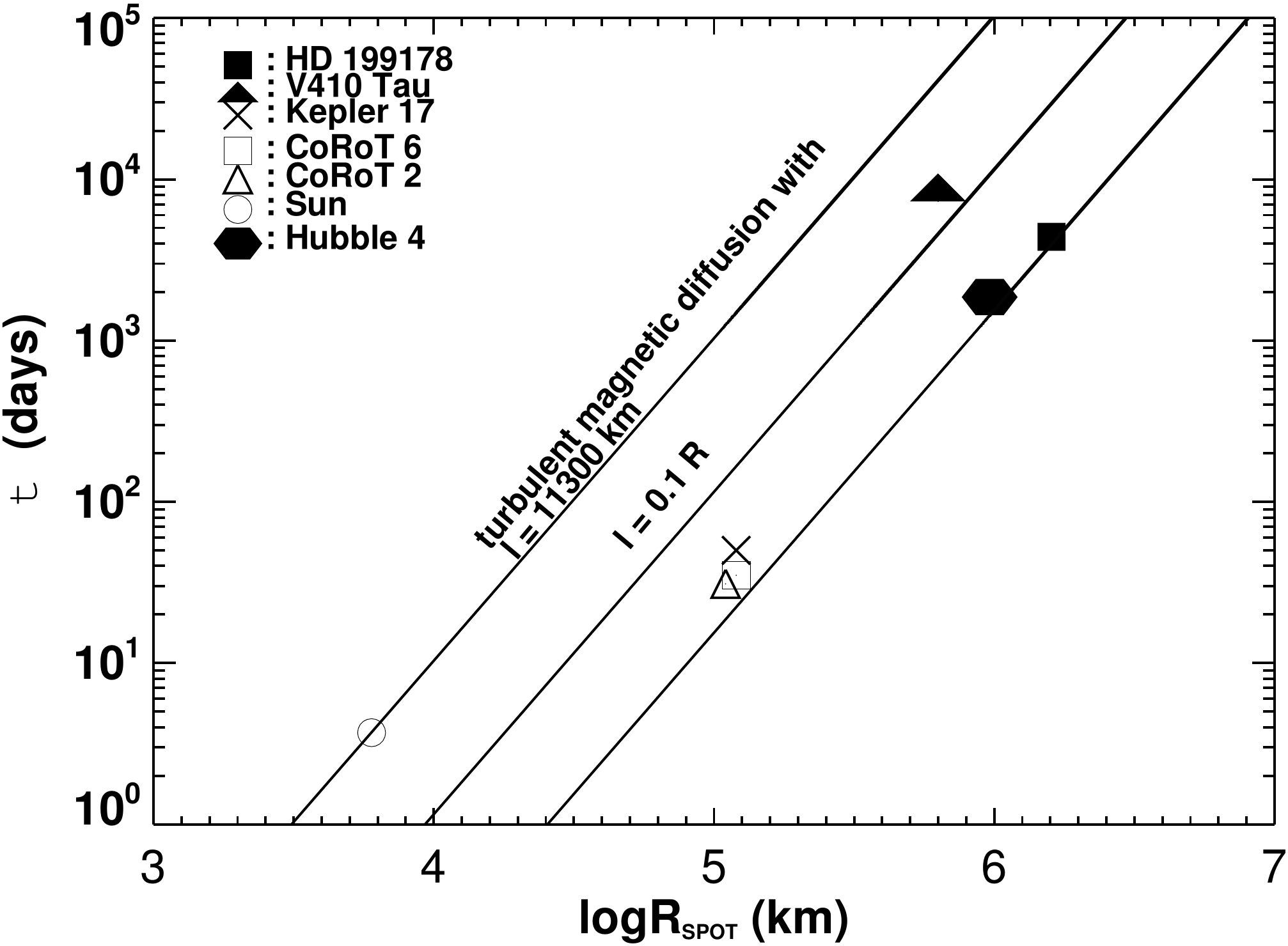}
	\caption{Observed spot durations and GW Law models for varying super-granule sizes \citep{bradshaw_sunspot_2014}. Hexagon indicates the spot on Hubble 4 A, as it would appear in their original plot.}
	\label{fig:hartigan}
\end{figure}

One of the initial motivations for this study was to determine how long an
individual spot or spot group on a TTS could last.  This has bearing
on any study that attempts to use RVs or photometric variability
to detect planetary mass companions.  In the case of Hubble 4, there is strong evidence 
that a large spot or spot group persisted in the same location on
the star for at least 5 years, which corresponds to $\sim 1200$ stellar
rotations.  This is much longer than spots last on the Sun; however, as
discussed by \citet{bradshaw_sunspot_2014}, such a long lived spot on a
TTS is not unexpected based on simple theoretical arguments.  These authors
examine the role of turbulent and magnetic diffusion in the decay of
spots and show that for low gravity stars with strong magnetic fields and
large spots (as is the case of TTSs), the decay time can be several years.
In order to place Hubble 4 within this context, we must determine the spot's 
physical size.  Assuming an angular radius of 36$^\circ$ from the spot 
modelling described above and a stellar radius of 2.1 R$_\odot$, the spot has a 
physical radius of $8.77 \times 10^5$ km.  With a lower limit to the
spot lifetime of 5.1 years, we can re-create Figure 1 of 
\citet{bradshaw_sunspot_2014}, placing Hubble 4 in the diagram as well.  
This is shown in Figure \ref{fig:hartigan}, where it is seen that Hubble 4 
lies close to HD 199178 and V410 Tau \citep{stelzer_weak-line_2003} and sits almost on the expected relation
with a supergranule size of 0.5 R$_\odot$.  The first order explanation
from this type of analysis is that large starspots live longer than smaller
spots as is expected if a diffusion process governs their lifetimes.
\citet{bradshaw_sunspot_2014} suggest that the solar and stellar data taken
together support the idea that turbulent-driven magnetic diffusion governs
the lifetimes of both sunspots and starspots and that supergranules
represent the maximum scale size for this diffusion. 

\section{The Behavior of H\texorpdfstring{$\alpha$}{alpha} and Two Serendipitous Flares} \label{sec:flare}

One of the characteristic properties of magnetically active stars is
chromospheric emisssion in the H$\alpha$ line, and WTTSs are no 
exception to this.  On the Sun, chromospheric H$\alpha$ emission (really a filling
in of the deep photospheric absorption line) is strongest in
plages, which are most prominent in solar active regions.  On active, late type stars of K and M spectral types, 
strong chromospheric activity produces H$\alpha$ lines in emission above the local
continuum.
The level of this chromospheric H$\alpha$ emission can vary on rotational 
and longer timescales as the coverage of solar plage-like structures visible 
on the stellar surface changes, but it can also change dramatically on short
timescales as the result of stellar flares.

Hydrogen lines can also be powerful probes of mass loss from gaseous exoplanets in close orbits around their host stars, so called "hot Jupiters."
\citet{vidal-madjar_extended_2003} were the first to use UV Ly-$\alpha$ observations
to detect mass outflow from a hot Jupiter; this has now been done
for several of these objects \citep[e.g.,][]{ehrenreich_hint_2012}.  While most
Hydrogen line observations of hot Jupiter mass loss have been done
with Ly-$\alpha$, H$\alpha$ has also been used to study outflows 
from hot Jupiters \citep[e.g.,][]{cauley_evidence_2017, chen_detection_2020}.  H$\alpha$
emission from a young ($\sim 2$ Myr), evaporating hot Jupiter has
also been reported \citep{johns-krull_h_2016}.  The
authors find excess H$\alpha$ emission in the WTTS PTFO 8-8695 which seems 
to move in velocity in phase with the period of a suspected planetary 
companion. The most likely explanation is that the emission comes from the 
companion. However, as described in \citet{johns-krull_h_2016}, another 
possible source is stellar activity from the host star itself. Looking at 
the H$\alpha$ profile variability of other WTTSs for similar effects will
help determine the cause of the variations observed in PTFO8-8695. 
Specifically, we wish to see if the type of H$\alpha$ variability seen
in PTFO8-8695 (an excess emission component, nearly as strong as the central 
stellar component, observed to move from one side of the main line to the 
other) is commonly observed on other WTTSs.  

\begin{figure}
    \centering
	\includegraphics[angle=90,width= 0.85\linewidth]{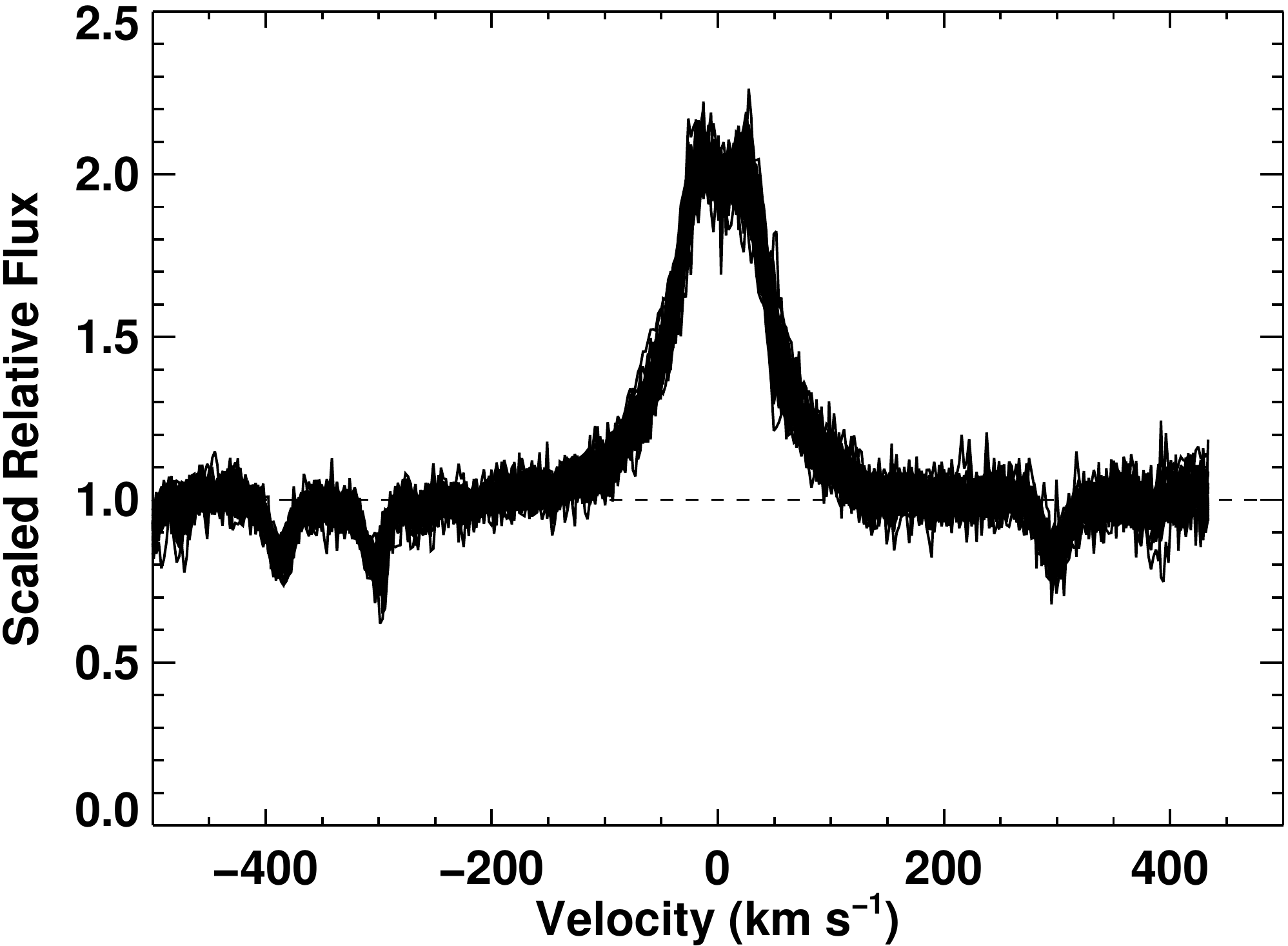}
	\caption{Overplotted H$\alpha$ emission lines from the 64 Hubble 4 system observations which do not contain a flare.}
	\label{fig:halphas}
\end{figure}

We examined the H$\alpha$ line profile variations of Hubble 4.
All but two of the line profiles display essentially the same shape,
with variations appearing primarily as a scaling up or down of the whole
profile together.  This can be seen in Figure \ref{fig:halphas} where
we have scaled each profile so that the central 0.9 \AA\ of the line has
the same mean flux; none of the displayed
profiles show a strong excess component out to $\sim 200$ 
km s$^{-1}$ on one side or the other as is seen in PTFO8-8695.  Thus,
the behavior seen in PTFO8-8695 does not appear to be present in Hubble 4.

\begin{figure}
    \centering
	\includegraphics[angle = 90, width= 0.85\linewidth]{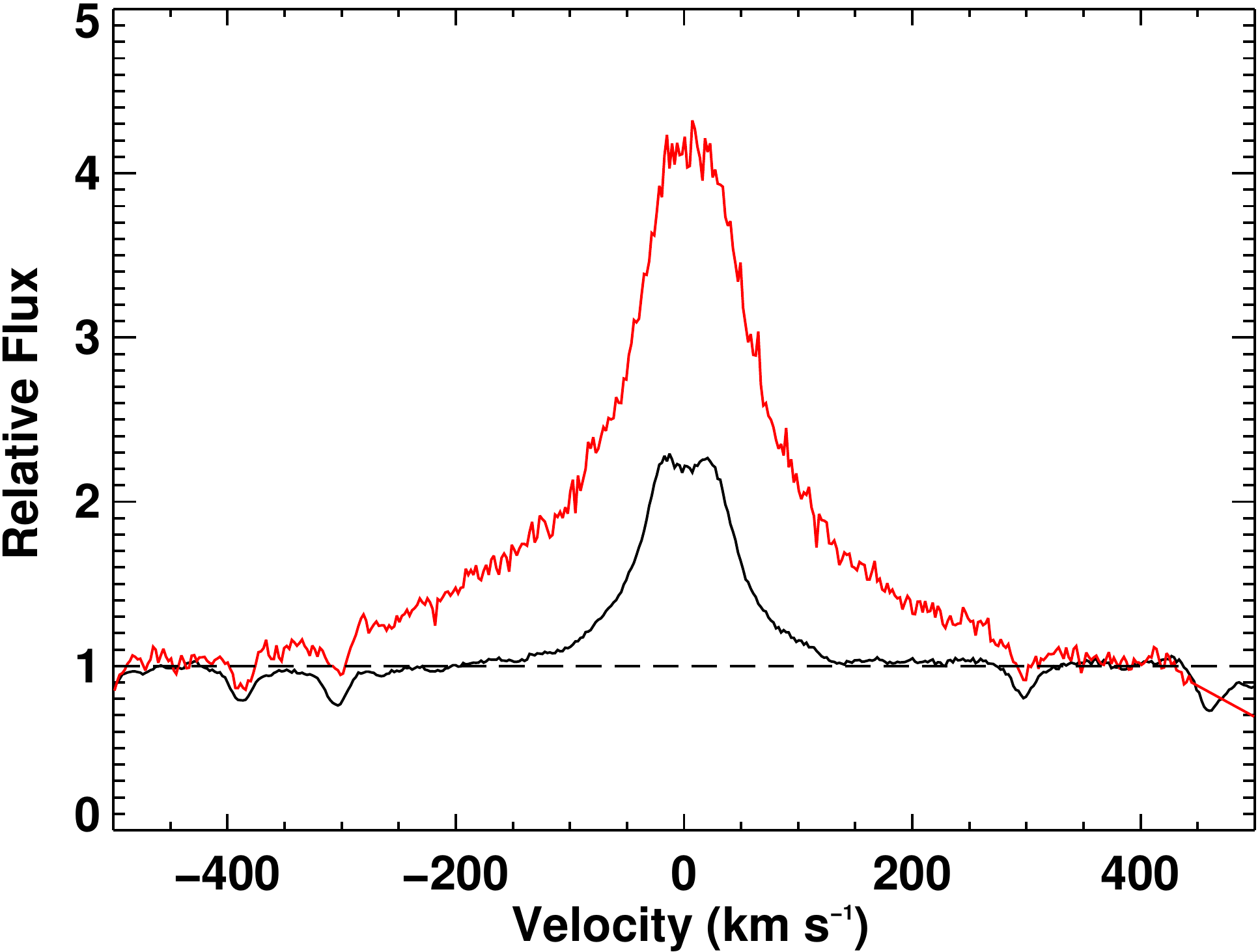}
	\caption{Relative flux of the $H\alpha$ line for the very large
	flare (upper profile in red) compared to the average $H_\alpha$ line
	profile in Hubble 4 for the UT 02:41:54.26 Feb 13, 2007 flare. Dashed line at 1.0 for reference to continuum.}
	\label{fig:flare}
\end{figure}

We did, however, appear to catch two flares on Hubble 4, with one of them
appearing to be quite powerful.  The stronger flare was seen in the
observation taken at UT 02:41:54.26 on February 13, 2007 (JD 2454144.62).
Figure \ref{fig:flare} shows the relative flux of the H$\alpha$ line during 
this observation, as well as the average across all the other non-flaring
observations.  The strong increase in emission with a roughly symmetric,
substantial increase in the line broadening is a common feature of
H$\alpha$ emission during flares on dMe stars \citep[e.g.,][]{honda_time-resolved_2018, vida_quest_2019}.
The measured equivalent width of the line was 13.43 \AA\ during the flare, 
whereas the equivalent width of the average line is 2.99 \AA, indicating an
increase in the H$\alpha$ emission by a factor of $\sim 4.5$ during the flare
averaged over the 1800 seconds of this exposure.  Assuming an approximate
$T_{eff} = 4500$ K for the primary star and again using the BT-NextGen atmosphere
models, the photospheric continuum spectral flux density at H$\alpha$ is 
$\sim 2.3\times 10^{6}$ erg $\text{cm}^{-2}\text{s}^{-1}\mbox{\AA}^{-1}$.
Using a radius of 2.1 R$_\odot$ for the primary star and the measured increase
in the equivalent width relative to the average for H$\alpha$, we calculated
that this flare released a total of $\sim 1.2 \times 10^{34}$ erg in the
H$\alpha$ line alone during the portion of the flare we recorded.  To estimate the total energy released in the flare during this time,
we assumed the energy in H$\alpha$ divided by the total U-band energy is
$\sim 0.06$ \citep[taken as a mean from flares studied in][]{kowalski_time-resolved_2012, kowalski_time-resolved_2013}, and that the U-band energy divided by the
total radiated energy is $\sim 0.11$ \citep{osten_connecting_2015}.  We then
calculated a total flare energy of $\sim 1.8 \times 10^{36}$ erg, significantly stronger than flare energies typically seen in dMe stars, $\sim 10^{34}$ erg captured across the 
entire visible range \citep{doyle_probing_2019}. This is also $\sim$ 10 times more energetic than flares reported on BP Tau \citep{gullbring_activity_1996}. While our
calculation of the flare energy is approximate, it is also a lower limit because
we may not have captured the entire flare in our exposure, and we only
consider the primary star when estimating the continuum flux density recorded.
The secondary star has a projected separation of $\sim 65$ mas and is therefore
well within our 1.2$^{\prime\prime}$ slit and contributes its flux to the
observed continuum.

\begin{figure}
    \centering
	\includegraphics[angle = 90, width= 0.85\linewidth]{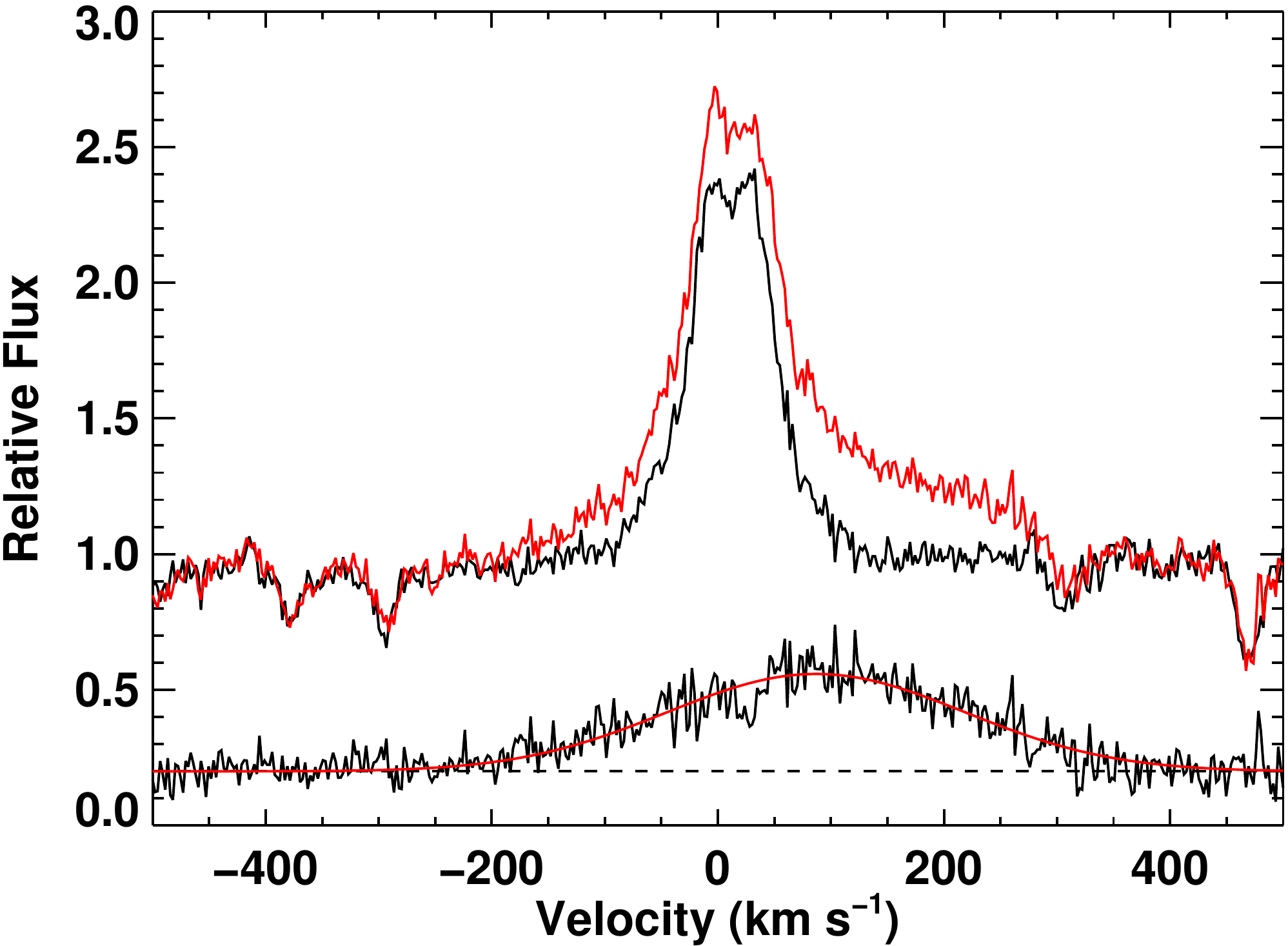}
	\caption{Relative flux of the H$\alpha$ line for the UT 11:10:50.6 November 17, 2018 flare
	(upper red profile) on Hubble 4 compared to the non-flaring H$\alpha$ line
	profile (upper black profile) observed earlier the same night. The difference of
	the flaring minus the non-flaring profile (offset upward by 0.2 for clarity)
	is shown at the bottom (lower black profile) with a Gaussian fit (lower red, smooth 
	line) overplotted. The dashed line provides the zero-point reference for the
	difference profile.}
	\label{fig:flare2}
\end{figure}

A second flare was seen in the observation taken at UT 11:10:50.6 on November 17,
2018 (JD 2458440.95).  On most nights during this run, we observed Hubble 4
twice: once at the beginning of the night and once at the end of the night.
Figure \ref{fig:flare2} shows the relative flux of the H$\alpha$ line during 
the flare observation compared to the spectrum of Hubble 4 taken 7.3 hours
earlier that same night when the star was not flaring.  Again, we see an overall increase in the 
H$\alpha$ line strength, including excess broadening in the wings.  However,
this time the observed excess is markedly asymmetric, with emission
extending further to the red than to the blue.  These types of asymmetric
increases in emission are also commonly observed in flares from dMe stars
\citep{honda_time-resolved_2018, vida_quest_2019} .  Simply subtracting the pre-flare
profile from the flaring profile results in a profile that is very
Gaussian in shape and displays a significant redshift (Figure \ref{fig:flare2}).
Redshifted emission components are often observed in both solar \citep[e.g.,][]{can90}
and stellar \citep[e.g.,][]{rice2011} flares.  The redshifts are usually interpreted
in terms of chromospheric evaporation and condensation that results from the intense
heating of the upper chromosphere by downward propagating beams of electrons produced
in the reconnection event which initiated the flare \citep[][]{fisher1985a,fisher1985b}.
Fitting a Gaussian to the
difference gives a velocity shift of +87 km s$^{-1}$ and a FWHM of 
310 km s$^{-1}$.  The equivalent width of the pre-flare H$\alpha$ profile is
3.02 \AA, and the equivalent width of the flaring profile is 5.43 \AA.
The exposure time for the flare observation is again 1800 s.  Using the
same method as above to estimate the minimum total energy in this flare
results in a value of $4.1 \times 10^{35}$ erg, another very strong flare
when compared to those typical of dMe stars.

\section{Conclusion} \label{sec:conclusion}

We have analyzed RV data taken over a total of 14 years for the WTTS Hubble 4 and combined
this with an analysis of HST imaging of the system.  The RV variations
show clear contributions from both the 9.3 year orbital motion of the binary system
as well as a clear $\sim$1.5 day signal that is presumably the rotation period of the brighter component of the binary.  The HST imaging data allowed us to estimate
the separation, position angle, and flux ratio of the component stars.  The
separation and position angle are consistent with the orbit determination
presented by \citet{galli_gouldtextquotesingles_2018}.  The flux ratio between the two stars
(which affects the RV signal produced by the blended lines from the orbiting pair, and is therefore constrained by the measured RV signal) is consistent with the masses determined from
\citet{galli_gouldtextquotesingles_2018} and the 2 Myr isochrones from \citet{baraffe_new_2015} if we take into account that the primary is substantially spotted.  

Subtracting the orbital contribution to the RV variations, we studied the
spot induced RV variations more closely.  Our analysis shows that a
large spot or spot group on Hubble 4 maintained its basic size and location
on the surface for a minimum of 5 years.  Further, the size of this spot
is consistent with the spot coverage needed to reconcile the imaging
and orbital RV analysis of the flux ratio between the two components.
Hubble 4's long lived spot fits well with the turbulent diffusion analysis
of spot sizes and lifetimes presented by \citet{bradshaw_sunspot_2014}.
These authors show how spot sizes and lifetimes can be used to infer the
size of supergranulation cells on stars, allowing studies such as the one
we have presented here to inform stellar dynamo and flux emergence models.
Equally important are the implications of our work
for planet searches around very young stars.  The primary planet search
methods (RV and transit) must contend with the effects of starspots 
which can produce astrophysical signals that mimic those of planets.
The extremely long lifetime and large amplitude variations of coherent spots
or spot structures on young stars can make it more difficult to search for planets around these stars. On the other hand, such stable structures offer possibilities to explore how to remove these signals using other measures of spot contributions, such as bisector span variations, and search for lower amplitude signals from low mass companions \citep[e.g.,][]{donati_hot_2017}.  It is therefore important to use multiple 
criteria to rule out the potential effects 

\acknowledgements
We thank the anonymous referee for comments that improved the manuscript.
This work was made possible in part by support from a Dessler Scholarship Grant through the Department of Physics and Astronomy at Rice University
during the summer of 2017.  Additional support for this research has been provided by the NSF through grant number AST-2009197 made to Rice University.  Partial support for L.P. was provided by NASA Exoplanet Research Program grant 80-NSSC19K-0289 and by the SIM Young Planets Key Project.  This research was based in part on observations made with the NASA/ESA Hubble Space Telescope, obtained from the data archive at the Space Telescope Science Institute. STScI is operated by the Association of Universities for Research in Astronomy, Inc. under NASA contract NAS 5-26555.  This research has made use of NASA's Astrophysics Data System Bibliographic Services.  This research has made use of the SIMBAD database,
operated at CDS, Strasbourg, France.

\bibliography{Hubble4Paper}{}

\begin{thebibliography}{}
\expandafter\ifx\csname natexlab\endcsname\relax\def\natexlab#1{#1}\fi
\providecommand{\url}[1]{\href{#1}{#1}}
\providecommand{\dodoi}[1]{doi:~\href{http://doi.org/#1}{\nolinkurl{#1}}}
\providecommand{\doeprint}[1]{\href{http://ascl.net/#1}{\nolinkurl{http://ascl.net/#1}}}
\providecommand{\doarXiv}[1]{\href{https://arxiv.org/abs/#1}{\nolinkurl{https://arxiv.org/abs/#1}}}

\bibitem[{Allard {et~al.}(2012)Allard, Homeier, \&
  Freytag}]{allard_models_2012}
Allard, F., Homeier, D., \& Freytag, B. 2012, Philosophical Transactions of the
  Royal Society A: Mathematical, Physical and Engineering Sciences, 370, 2765,
  \dodoi{10.1098/rsta.2011.0269}

\bibitem[{Asplund {et~al.}(2009)Asplund, Grevesse, Sauval, \&
  Scott}]{asplund_chemical_2009}
Asplund, M., Grevesse, N., Sauval, A.~J., \& Scott, P. 2009, Annual Review of
  Astronomy and Astrophysics, 47, 481,
  \dodoi{10.1146/annurev.astro.46.060407.145222}

\bibitem[{Baraffe {et~al.}(2015)Baraffe, Homeier, Allard, \&
  Chabrier}]{baraffe_new_2015}
Baraffe, I., Homeier, D., Allard, F., \& Chabrier, G. 2015, Astronomy \&
  Astrophysics, 577, A42, \dodoi{10.1051/0004-6361/201425481}

\bibitem[{Bouvier {et~al.}(1993)Bouvier, Cabrit, Fernandez, Martin, \&
  Matthews}]{bouvier_coyotes-i_1993}
Bouvier, J., Cabrit, S., Fernandez, M., Martin, E.~L., \& Matthews, J.~M. 1993,
  Astronomy and Astrophysics, 272, 176.
\newblock \url{http://adsabs.harvard.edu/abs/1993A%26A...272..176B}

\bibitem[{Boyajian {et~al.}(2016)Boyajian, LaCourse, Rappaport, Fabrycky,
  Fischer, Gandolfi, Kennedy, Korhonen, Liu, Moor, Olah, Vida, Wyatt, Best,
  Brewer, Ciesla, Csák, Deeg, Dupuy, Handler, Heng, Howell, Ishikawa, Kovács,
  Kozakis, Kriskovics, Lehtinen, Lintott, Lynn, Nespral, Nikbakhsh, Schawinski,
  Schmitt, Smith, Szabo, Szabo, Viuho, Wang, Weiksnar, Bosch, Connors, Goodman,
  Green, Hoekstra, Jebson, Jek, Omohundro, Schwengeler, \&
  Szewczyk}]{boyajian_planet_2016}
Boyajian, T.~S., LaCourse, D.~M., Rappaport, S.~A., {et~al.} 2016, Monthly
  Notices of the Royal Astronomical Society, 457, 3988,
  \dodoi{10.1093/mnras/stw218}

\bibitem[{Bradshaw \& Hartigan(2014)}]{bradshaw_sunspot_2014}
Bradshaw, S.~J., \& Hartigan, P. 2014, The Astrophysical Journal, 795, 79,
  \dodoi{10.1088/0004-637X/795/1/79}

\bibitem[{Canfield {et~al.}(1990)Canfield, Penn, Wulser, \& Kiplinger}]{can90}
Canfield, R.~C., Penn, M.~J., Wulser, J.-P., \& Kiplinger, A.~L. 1990,
  {\textbackslash}apj, 363, 318, \dodoi{10.1086/169345}

\bibitem[{Cauley {et~al.}(2017)Cauley, Redfield, \&
  Jensen}]{cauley_evidence_2017}
Cauley, P.~W., Redfield, S., \& Jensen, A.~G. 2017, The Astronomical Journal,
  153, 185, \dodoi{10.3847/1538-3881/aa64d3}

\bibitem[{Chen {et~al.}(2020)Chen, Casasayas-Barris, Pallé, Yan, Stangret,
  Cegla, Allart, \& Lovis}]{chen_detection_2020}
Chen, G., Casasayas-Barris, N., Pallé, E., {et~al.} 2020, Astronomy and
  Astrophysics, 635, A171, \dodoi{10.1051/0004-6361/201936986}

\bibitem[{Chen \& Johns-Krull(2013)}]{chen_spectropolarimetry_2013}
Chen, W., \& Johns-Krull, C.~M. 2013, The Astrophysical Journal, 776, 113,
  \dodoi{10.1088/0004-637X/776/2/113}

\bibitem[{Choi \& Herbst(1996)}]{choi_rotation_1996}
Choi, P.~I., \& Herbst, W. 1996, The Astronomical Journal, 111, 283,
  \dodoi{10.1086/117780}

\bibitem[{{Czekala} {et~al.}(2019){Czekala}, {Chiang}, {Andrews}, {Jensen},
  {Torres}, {Wilner}, {Stassun}, \& {Macintosh}}]{Cze2019}
{Czekala}, I., {Chiang}, E., {Andrews}, S.~M., {et~al.} 2019, \apj, 883, 22,
  \dodoi{10.3847/1538-4357/ab287b}

\bibitem[{Donati \& Cameron(1997)}]{donati_differential_1997}
Donati, J.-F., \& Cameron, A.~C. 1997, Monthly Notices of the Royal
  Astronomical Society, 291, 1, \dodoi{10.1093/mnras/291.1.1}

\bibitem[{Donati {et~al.}(2017)Donati, Yu, Moutou, Cameron, Malo, Grankin,
  Hébrard, Hussain, Vidotto, Alencar, Haywood, Bouvier, Petit, Takami,
  Herczeg, Gregory, Jardine, Morin, \& {MaTYSSE
  Collaboration}}]{donati_hot_2017}
Donati, J.-F., Yu, L., Moutou, C., {et~al.} 2017, Monthly Notices of the Royal
  Astronomical Society, 465, 3343, \dodoi{10.1093/mnras/stw2904}

\bibitem[{Doyle {et~al.}(2019)Doyle, Ramsay, Doyle, \& Wu}]{doyle_probing_2019}
Doyle, L., Ramsay, G., Doyle, J.~G., \& Wu, K. 2019, Monthly Notices of the
  Royal Astronomical Society, 489, 437, \dodoi{10.1093/mnras/stz2205}

\bibitem[{Ehrenreich {et~al.}(2012)Ehrenreich, Bourrier, Bonfils, Etangs,
  Hébrard, Sing, Wheatley, Vidal-Madjar, Delfosse, Udry, Forveille, \&
  Moutou}]{ehrenreich_hint_2012}
Ehrenreich, D., Bourrier, V., Bonfils, X., {et~al.} 2012, Astronomy \&
  Astrophysics, 547, A18, \dodoi{10.1051/0004-6361/201219981}

\bibitem[{Fischer {et~al.}(2014)Fischer, Marcy, \&
  Spronck}]{fischer_twenty-five_2014}
Fischer, D.~A., Marcy, G.~W., \& Spronck, J. F.~P. 2014, The Astrophysical
  Journal Supplement Series, 210, 5, \dodoi{10.1088/0067-0049/210/1/5}

\bibitem[{Fisher {et~al.}(1985{\natexlab{a}})Fisher, Canfield, \&
  McClymont}]{fisher1985a}
Fisher, G.~H., Canfield, R.~C., \& McClymont, A.~N. 1985{\natexlab{a}},
  {\textbackslash}apj, 289, 434, \dodoi{10.1086/162903}

\bibitem[{Fisher {et~al.}(1985{\natexlab{b}})Fisher, Canfield, \&
  McClymont}]{fisher1985b}
---. 1985{\natexlab{b}}, {\textbackslash}apj, 289, 425, \dodoi{10.1086/162902}

\bibitem[{Galli {et~al.}(2018)Galli, Loinard, Ortiz-Léon, Kounkel, Dzib,
  Mioduszewski, Rodríguez, Hartmann, Teixeira, Torres, Rivera, Boden, II,
  Briceño, Tobin, \& Heyer}]{galli_gouldtextquotesingles_2018}
Galli, P. A.~B., Loinard, L., Ortiz-Léon, G.~N., {et~al.} 2018, The
  Astrophysical Journal, 859, 33, \dodoi{10.3847/1538-4357/aabf91}

\bibitem[{Gray(2005)}]{gray_observation_2005}
Gray, D.~F. 2005, The {Observation} and {Analysis} of {Stellar} {Photospheres},
  \dodoi{10.1017/CBO9781316036570}

\bibitem[{Gullbring {et~al.}(1996)Gullbring, Barwig, Chen, Gahm, \&
  Bao}]{gullbring_activity_1996}
Gullbring, E., Barwig, H., Chen, P.~S., Gahm, G.~F., \& Bao, M.~X. 1996,
  Astronomy and Astrophysics, 307, 791.
\newblock \url{http://adsabs.harvard.edu/abs/1996A%26A...307..791G}

\bibitem[{Hartig(2008)}]{hartig_wfc3_2008}
Hartig, G.~F. 2008, STScI ISR WFC3-2008-44, 10, 100

\bibitem[{Hartmann {et~al.}(1986)Hartmann, Hewett, Stahler, \&
  Mathieu}]{hartmann_rotational_1986}
Hartmann, L., Hewett, R., Stahler, S., \& Mathieu, R.~D. 1986, The
  Astrophysical Journal, 309, 275, \dodoi{10.1086/164599}

\bibitem[{Hatzes(1995)}]{hatzes_doppler_1995}
Hatzes, A.~P. 1995, The Astrophysical Journal, 451, 784, \dodoi{10.1086/176265}

\bibitem[{Hauschildt {et~al.}(1999)Hauschildt, Allard, \&
  Baron}]{hauschildt_nextgen_1999}
Hauschildt, P.~H., Allard, F., \& Baron, E. 1999, The Astrophysical Journal,
  512, 377, \dodoi{10.1086/306745}

\bibitem[{Hinkle {et~al.}(2000)Hinkle, Joyce, Sharp, \&
  Valenti}]{hinkle_phoenix_2000}
Hinkle, K.~H., Joyce, R.~R., Sharp, N., \& Valenti, J.~A. 2000, in Optical and
  {IR} {Telescope} {Instrumentation} and {Detectors}, Vol. 4008 (International
  Society for Optics and Photonics), 720--728, \dodoi{10.1117/12.395529}

\bibitem[{Honda {et~al.}(2018)Honda, Notsu, Namekata, Notsu, Maehara, Ikuta,
  Nogami, \& Shibata}]{honda_time-resolved_2018}
Honda, S., Notsu, Y., Namekata, K., {et~al.} 2018, Publications of the
  Astronomical Society of Japan, 70, \dodoi{10.1093/pasj/psy055}

\bibitem[{{Horne} \& {Baliunas}(1986)}]{horne1986}
{Horne}, J.~H., \& {Baliunas}, S.~L. 1986, \apj, 302, 757,
  \dodoi{10.1086/164037}

\bibitem[{Houk \& Swift(1999)}]{houk_michigan_1999}
Houk, N., \& Swift, C. 1999, 05, 0.
\newblock \url{http://adsabs.harvard.edu/abs/1999MSS...C05....0H}

\bibitem[{Huerta {et~al.}(2008)Huerta, Johns-Krull, Prato, Hartigan, \&
  Jaffe}]{huerta_starspot-induced_2008}
Huerta, M., Johns-Krull, C.~M., Prato, L., Hartigan, P., \& Jaffe, D.~T. 2008,
  The Astrophysical Journal, 678, 472, \dodoi{10.1086/526415}

\bibitem[{Johns-Krull \& Herczeg(2007)}]{johns-krull_how_2007}
Johns-Krull, C.~M., \& Herczeg, G.~J. 2007, The Astrophysical Journal, 655,
  345, \dodoi{10.1086/508770}

\bibitem[{Johns-Krull {et~al.}(2004)Johns-Krull, Valenti, \&
  Saar}]{johns-krull_testing_2004}
Johns-Krull, C.~M., Valenti, J.~A., \& Saar, S.~H. 2004, The Astrophysical
  Journal, 617, 1204, \dodoi{10.1086/425652}

\bibitem[{Johns-Krull {et~al.}(2016)Johns-Krull, Prato, McLane, Ciardi, van
  Eyken, Chen, Stauffer, Beichman, Frazier, Boden, Morales-Calderón, \&
  Rebull}]{johns-krull_h_2016}
Johns-Krull, C.~M., Prato, L., McLane, J.~N., {et~al.} 2016, The Astrophysical
  Journal, 830, 15, \dodoi{10.3847/0004-637X/830/1/15}

\bibitem[{Kalirai {et~al.}(2009)Kalirai, MacKenty, Rajan, Baggett, Bohlin,
  Brown, Deustua, Kimble, Riess, \& Sabbi}]{kalirai_wfc3_2009}
Kalirai, J.~S., MacKenty, J., Rajan, A., {et~al.} 2009, Space Telescope WFC
  Instrument Science Report.
\newblock \url{http://adsabs.harvard.edu/abs/2009wfc..rept...31K}

\bibitem[{Kowalski(2012)}]{kowalski_time-resolved_2012}
Kowalski, A.~F. 2012, PhD thesis.
\newblock \url{http://adsabs.harvard.edu/abs/2012PhDT.......179K}

\bibitem[{Kowalski {et~al.}(2013)Kowalski, Hawley, Wisniewski, Osten, Hilton,
  Holtzman, Schmidt, \& Davenport}]{kowalski_time-resolved_2013}
Kowalski, A.~F., Hawley, S.~L., Wisniewski, J.~P., {et~al.} 2013, The
  Astrophysical Journal Supplement Series, 207, 15,
  \dodoi{10.1088/0067-0049/207/1/15}

\bibitem[{Kraus {et~al.}(2011)Kraus, Ireland, Martinache, \&
  Hillenbrand}]{kraus_mapping_2011}
Kraus, A.~L., Ireland, M.~J., Martinache, F., \& Hillenbrand, L.~A. 2011, The
  Astrophysical Journal, 731, 8, \dodoi{10.1088/0004-637X/731/1/8}

\bibitem[{{Kurtovic} {et~al.}(2018){Kurtovic}, {P{\'e}rez}, {Benisty}, {Zhu},
  {Zhang}, {Huang}, {Andrews}, {Dullemond}, {Isella}, {Bai}, {Carpenter},
  {Guzm{\'a}n}, {Ricci}, \& {Wilner}}]{Kurt2018}
{Kurtovic}, N.~T., {P{\'e}rez}, L.~M., {Benisty}, M., {et~al.} 2018, \apjl,
  869, L44, \dodoi{10.3847/2041-8213/aaf746}

\bibitem[{Mahmud {et~al.}(2011)Mahmud, Crockett, Johns-Krull, Prato, Hartigan,
  Jaffe, \& Beichman}]{mahmud_starspot-induced_2011}
Mahmud, N.~I., Crockett, C.~J., Johns-Krull, C.~M., {et~al.} 2011, The
  Astrophysical Journal, 736, 123, \dodoi{10.1088/0004-637X/736/2/123}

\bibitem[{Mann {et~al.}(2016)Mann, Newton, Rizzuto, Irwin, Feiden, Gaidos,
  Mace, Kraus, James, Ansdell, Charbonneau, Covey, Ireland, Jaffe, Johnson,
  Kidder, \& Vanderburg}]{mann_zodiacal_2016}
Mann, A.~W., Newton, E.~R., Rizzuto, A.~C., {et~al.} 2016, The Astronomical
  Journal, 152, 61, \dodoi{10.3847/0004-6256/152/3/61}

\bibitem[{Norton {et~al.}(2007)Norton, Wheatley, West, Haswell, Street,
  Cameron, Christian, Clarkson, Enoch, Gallaway, Hellier, Horne, Irwin, Kane,
  Lister, Nicholas, Parley, Pollacco, Ryans, Skillen, \&
  Wilson}]{norton_new_2007}
Norton, A.~J., Wheatley, P.~J., West, R.~G., {et~al.} 2007, Astronomy \&
  Astrophysics, 467, 785, \dodoi{10.1051/0004-6361:20077084}

\bibitem[{Osten \& Wolk(2015)}]{osten_connecting_2015}
Osten, R.~A., \& Wolk, S.~J. 2015, The Astrophysical Journal, 809, 79,
  \dodoi{10.1088/0004-637X/809/1/79}

\bibitem[{Pecaut \& Mamajek(2013)}]{pecaut_intrinsic_2013}
Pecaut, M.~J., \& Mamajek, E.~E. 2013, The Astrophysical Journal Supplement
  Series, 208, 9, \dodoi{10.1088/0067-0049/208/1/9}

\bibitem[{Piskunov {et~al.}(1995)Piskunov, Kupka, Ryabchikova, Weiss, \&
  Jeffery}]{piskunov_vald_1995}
Piskunov, N.~E., Kupka, F., Ryabchikova, T.~A., Weiss, W.~W., \& Jeffery, C.~S.
  1995, Astronomy and Astrophysics Supplement Series, 112, 525.
\newblock \url{http://adsabs.harvard.edu/abs/1995A%26AS..112..525P}

\bibitem[{Rice {et~al.}(2011)Rice, Strassmeier, \& Kopf}]{rice2011}
Rice, J.~B., Strassmeier, K.~G., \& Kopf, M. 2011, {\textbackslash}apj, 728,
  69, \dodoi{10.1088/0004-637X/728/1/69}

\bibitem[{Rizzuto {et~al.}(2020)Rizzuto, Dupuy, Ireland, \&
  Kraus}]{rizzuto_dynamical_2020}
Rizzuto, A.~C., Dupuy, T.~J., Ireland, M.~J., \& Kraus, A.~L. 2020, The
  Astrophysical Journal, 889, 175, \dodoi{10.3847/1538-4357/ab5aed}

\bibitem[{Robertson {et~al.}(2020)Robertson, Stefansson, Mahadevan, Endl,
  Cochran, Beard, Bender, Diddams, Duong, Ford, Fredrick, Halverson, Hearty,
  Holcomb, Juan, Kanodia, Lubin, Metcalf, Monson, Ninan, Palafoutas, Ramsey,
  Roy, Schwab, Terrien, \& Wright}]{robertson_persistent_2020}
Robertson, P., Stefansson, G., Mahadevan, S., {et~al.} 2020, The Astrophysical
  Journal, 897, 125, \dodoi{10.3847/1538-4357/ab989f}

\bibitem[{Ryabchikova {et~al.}(2015)Ryabchikova, Piskunov, Kurucz, Stempels,
  Heiter, Pakhomov, \& Barklem}]{ryabchikova_major_2015}
Ryabchikova, T., Piskunov, N., Kurucz, R.~L., {et~al.} 2015, Physica Scripta,
  90, 054005, \dodoi{10.1088/0031-8949/90/5/054005}

\bibitem[{Saar \& Donahue(1997)}]{saar_activity-related_1997}
Saar, S.~H., \& Donahue, R.~A. 1997, The Astrophysical Journal, 485, 319,
  \dodoi{10.1086/304392}

\bibitem[{Sabbi(2009)}]{sabbi_uvis_2009}
Sabbi, E. 2009, HST Proposal.
\newblock \url{http://adsabs.harvard.edu/abs/2009hst..prop11798S}

\bibitem[{Skinner(1993)}]{skinner_circularly_1993}
Skinner, S.~L. 1993, The Astrophysical Journal, 408, 660,
  \dodoi{10.1086/172626}

\bibitem[{Soderblom {et~al.}(1989)Soderblom, Pendleton, \&
  Pallavicini}]{soderblom_calibration_1989}
Soderblom, D.~R., Pendleton, J., \& Pallavicini, R. 1989, The Astronomical
  Journal, 97, 539, \dodoi{10.1086/115003}

\bibitem[{Soto \& Jenkins(2018)}]{soto_spectroscopic_2018}
Soto, M.~G., \& Jenkins, J.~S. 2018, Astronomy \& Astrophysics, 615, A76,
  \dodoi{10.1051/0004-6361/201731533}

\bibitem[{Stelzer {et~al.}(2003{\natexlab{a}})Stelzer, Fernández, Costa,
  Gameiro, Grankin, Henden, Guenther, Mohanty, Flaccomio, Burwitz,
  Jayawardhana, Predehl, \& Durisen}]{stelzer_weak-line_2003}
Stelzer, B., Fernández, M., Costa, V.~M., {et~al.} 2003{\natexlab{a}},
  Astronomy and Astrophysics, 411, 517, \dodoi{10.1051/0004-6361:20031414}

\bibitem[{Stelzer {et~al.}(2003{\natexlab{b}})Stelzer, Fernández, Costa,
  Gameiro, Grankin, Henden, Guenther, Mohanty, Flaccomio, Burwitz,
  Jayawardhana, Predehl, \& Durisen}]{stelzer_weak-line_2003-1}
---. 2003{\natexlab{b}}, Astronomy \& Astrophysics, 411, 517,
  \dodoi{10.1051/0004-6361:20031414}

\bibitem[{Tonry \& Davis(1979)}]{tonry_survey_1979}
Tonry, J., \& Davis, M. 1979, The Astronomical Journal, 84, 1511,
  \dodoi{10.1086/112569}

\bibitem[{Tull {et~al.}(1995)Tull, MacQueen, Sneden, \&
  Lambert}]{tull_high-resolution_1995}
Tull, R.~G., MacQueen, P.~J., Sneden, C., \& Lambert, D.~L. 1995, Publications
  of the Astronomical Society of the Pacific, 107, 251, \dodoi{10.1086/133548}

\bibitem[{Valenti(1994)}]{valenti_photospheric_1994}
Valenti, J.~A. 1994, PhD thesis, University of California, Berkeley.
\newblock \url{http://adsabs.harvard.edu/abs/1994PhDT........16V}

\bibitem[{Valenti \& Fischer(2005)}]{valenti_vizier_2005}
Valenti, J.~A., \& Fischer, D.~A. 2005, VizieR Online Data Catalog, 215.
\newblock \url{http://adsabs.harvard.edu/abs/2005yCat..21590141V}

\bibitem[{Vida {et~al.}(2019)Vida, Leitzinger, Kriskovics, Seli, Odert,
  Kovács, Korhonen, \& Driel-Gesztelyi}]{vida_quest_2019}
Vida, K., Leitzinger, M., Kriskovics, L., {et~al.} 2019, Astronomy \&
  Astrophysics, 623, A49, \dodoi{10.1051/0004-6361/201834264}

\bibitem[{Vidal-Madjar {et~al.}(2003)Vidal-Madjar, des Etangs, Désert,
  Ballester, Ferlet, Hébrard, \& Mayor}]{vidal-madjar_extended_2003}
Vidal-Madjar, A., des Etangs, A.~L., Désert, J.-M., {et~al.} 2003, Nature,
  422, 143, \dodoi{10.1038/nature01448}

\end{thebibliography}
\bibliographystyle{aasjournal}

\appendix 

\restartappendixnumbering
\section{RV and Bisector Span Measurements} \label{appendix:app1}

Table A1 presents the full set of RV and bisector span measurements obtained for
Hubble 4.

\startlongtable
\begin{deluxetable*}{cccccc}
	\tablecaption{Hubble 4 RV and Bisector Span Measurements\label{tab:h4_rvs}}
    
    \tablewidth{0pt}
	\tablehead{
		\colhead{Civil Date} & \colhead{Julian Date} & \colhead{RVs} & \colhead{$\sigma_{RV}$} 
		& \colhead{Bisector Spans} & \colhead{$\sigma_{bisector}$}\\
		\colhead{} & \colhead{(JD - 2450000)} & \colhead{($\text{kms}^{-1}$)} & \colhead{($\text{kms}^{-1}$)} & \colhead{($\text{kms}^{-1}$)} & \colhead{($\text{kms}^{-1}$)}
	}
	\startdata
03 Jan 2005 &             3373.810 &                 2.70 &                 0.28 &                 2.66 &                 0.73 \\
19 Nov 2005 &             3693.842 &                 2.31 &                 0.25 &                 2.73 &                 0.45 \\
20 Nov 2005 &             3694.750 &                 0.37 &                 0.20 &                 1.47 &                 0.54 \\
03 Feb 2006 &             3769.642 &                 2.58 &                 0.27 &                 2.86 &                 0.46 \\
04 Feb 2006 &             3770.631 &                 1.23 &                 0.26 &                 2.85 &                 0.36 \\
05 Feb 2006 &             3771.698 &                 0.97 &                 0.23 &                 2.61 &                 0.33 \\
08 Feb 2006 &             3774.733 &                 1.38 &                 0.25 &                 9.14 &                 1.01 \\
09 Feb 2006 &             3775.694 &                 2.18 &                 0.27 &                 3.17 &                 0.51 \\
08 Feb 2007 &             4139.599 &                 1.29 &                 0.29 &                 2.98 &                 0.27 \\
09 Feb 2007 &             4140.607 &                 3.18 &                 0.29 &                 4.25 &                 0.44 \\
10 Feb 2007 &             4141.601 &                 1.63 &                 0.25 &                 3.43 &                 0.27 \\
11 Feb 2007 &             4142.604 &                 1.62 &                 0.27 &                 3.54 &                 0.35 \\
12 Feb 2007 &             4143.723 &                 3.85 &                 0.31 &                 3.65 &                 0.48 \\
13 Feb 2007 &             4144.623 &                 1.69 &                 0.26 &                 4.11 &                 0.55 \\
20 Nov 2007 &             4424.819 &                 3.65 &                 0.23 &                 4.89 &                 0.45 \\
21 Nov 2007 &             4425.657 &                 1.59 &                 0.26 &                 3.16 &                 0.46 \\
26 Nov 2007 &             4430.750 &                 2.11 &                 0.28 &                 4.21 &                 0.43 \\
27 Nov 2007 &             4431.754 &                 1.65 &                 0.25 &                 3.97 &                 0.35 \\
28 Nov 2007 &             4432.713 &                 3.44 &                 0.33 &                 4.74 &                 0.50 \\
18 Nov 2008 &             4788.674 &                 0.01 &                 0.20 &                 1.03 &                 0.21 \\
19 Nov 2008 &             4789.674 &                 0.86 &                 0.19 &                 1.22 &                 0.20 \\
20 Nov 2008 &             4790.671 &                -1.65 &                 0.20 &                 0.67 &                 0.19 \\
22 Nov 2008 &             4792.667 &                 0.23 &                 0.22 &                 1.04 &                 0.29 \\
23 Nov 2008 &             4793.678 &                -1.14 &                 0.21 &                 0.74 &                 0.28 \\
24 Nov 2008 &             4794.664 &                 0.74 &                 0.20 &                 1.20 &                 0.20 \\
25 Nov 2008 &             4795.675 &                 0.01 &                 0.21 &                 0.43 &                 0.42 \\
11 Jan 2009 &             4842.619 &                -0.01 &                 0.24 &                 4.21 &                 0.57 \\
12 Jan 2009 &             4843.607 &                -0.81 &                 0.22 &                 1.25 &                 0.25 \\
13 Jan 2009 &             4844.593 &                -2.00 &                 0.21 &                 0.44 &                 0.20 \\
14 Jan 2009 &             4845.585 &                 0.66 &                 0.25 &                 1.36 &                 0.11 \\
15 Jan 2009 &             4846.587 &                -1.07 &                 0.24 &                 0.64 &                 0.25 \\
16 Jan 2009 &             4847.588 &                -1.24 &                 0.25 &                 0.28 &                 0.25 \\
17 Jan 2009 &             4848.592 &                 0.46 &                 0.23 &                 0.72 &                 0.23 \\
18 Jan 2009 &             4849.579 &                -1.66 &                 0.24 &                 0.06 &                 0.23 \\
24 Nov 2009 &             5159.765 &                -1.31 &                 0.21 &                 0.63 &                 0.34 \\
25 Nov 2009 &             5160.761 &                 0.51 &                 0.19 &                 1.01 &                 0.21 \\
26 Nov 2009 &             5161.729 &                -2.18 &                 0.19 &                 0.00 &                 0.00 \\
27 Nov 2009 &             5162.764 &                -0.54 &                 0.20 &                 0.42 &                 0.19 \\
28 Nov 2009 &             5163.848 &                 0.52 &                 0.21 &                 0.76 &                 0.40 \\
29 Nov 2009 &             5164.902 &                -2.75 &                 0.27 &                -0.81 &                 0.48 \\
02 Feb 2010 &             5229.581 &                -1.74 &                 0.28 &                 0.52 &                 0.23 \\
05 Feb 2010 &             5232.581 &                -1.67 &                 0.27 &                 1.00 &                 0.48 \\
06 Feb 2010 &             5233.564 &                 1.49 &                 0.28 &                 2.21 &                 0.19 \\
07 Feb 2010 &             5234.564 &                -1.13 &                 0.26 &                 1.17 &                 0.21 \\
08 Feb 2010 &             5235.564 &                -0.71 &                 0.27 &                 1.32 &                 0.29 \\
28 Nov 2017 &             8085.852 &                 1.53 &                 0.21 &                 3.77 &                 0.42 \\
30 Nov 2017 &             8087.930 &                 1.83 &                 0.22 &                 3.57 &                 0.37 \\
14 Nov 2018 &             8436.711 &                -1.70 &                 0.23 &                 0.20 &                 0.33 \\
15 Nov 2018 &             8437.703 &                -1.28 &                 0.23 &                -0.19 &                 0.27 \\
15 Nov 2018 &             8437.951 &                -2.28 &                 0.22 &                 0.14 &                 0.19 \\
16 Nov 2018 &             8438.695 &                -0.45 &                 0.22 &                 0.34 &                 0.14 \\
16 Nov 2018 &             8438.991 &                -0.93 &                 0.24 &                 0.47 &                 0.22 \\
17 Nov 2018 &             8439.677 &                -1.55 &                 0.25 &                 0.21 &                 0.41 \\
17 Nov 2018 &             8439.977 &                -1.26 &                 0.24 &                 0.18 &                 0.18 \\
18 Nov 2018 &             8440.688 &                -1.28 &                 0.25 &                -0.06 &                 0.28 \\
18 Nov 2018 &             8440.951 &                -2.10 &                 0.27 &                 0.31 &                 0.28 \\
27 Nov 2018 &             8449.987 &                -1.37 &                 0.24 &                 0.19 &                 0.55 \\
30 Nov 2018 &             8452.716 &                -0.75 &                 0.20 &                 0.39 &                 0.33 \\
02 Dec 2018 &             8454.977 &                -1.64 &                 0.23 &                -0.67 &                 0.19 \\
03 Dec 2018 &             8455.981 &                -0.85 &                 0.24 &                 0.97 &                 0.62 \\
20 Jan 2019 &             8503.649 &                -0.52 &                 0.27 &                 2.34 &                 0.27 \\
21 Jan 2019 &             8504.610 &                -1.35 &                 0.26 &                 3.14 &                 0.41 \\
22 Jan 2019 &             8505.590 &                -0.58 &                 0.28 &                 1.34 &                 0.24 \\
24 Jan 2019 &             8507.608 &                -1.54 &                 0.26 &                 0.12 &                 0.17 
     \enddata
\end{deluxetable*}

\section{Double-Line Spectroscopic Analysis for Fitting the Binary's RV Variations} \label{appendix:2line}

Hubble 4 is a binary with a well-determined orbit \citep{galli_gouldtextquotesingles_2018}
inclined to within $\sim 26^\circ$ of face on. The predicted peak-to-peak RV amplitude of each star is $\sim 8$ km s$^{-1}$ for the primary and $\sim 13.5$ km s$^{-1}$ for the secondary. The maximum predicted RV separation between the two components at the times we observed Hubble 4 is $\sim 11$ km s$^{-1}$, which is less than the $v$sin$i$ of $\sim14$ km s$^{-1}$ (14.0 km s$^{-1}$, see Appendix \ref{sec:vsini}).  As a result, the measurement of the orbital-induced RV signal for this system represents the average of the two stars' motion captured by the blended line profiles. These are mildly distorted as the result of the different continuum brightness
and intrinsic line depth of the two components at different epochs.  The CCF analysis of these profiles
measured apparent RV variations which encode some of the orbital information, but the measured
RV signal of these blended lines is not the true RV signal of either star.  We modeled in a more physically
motivated way the component RVs of the binary orbit to verify that we could
reproduce our measured RV variations in a self-consistent manner, while also verifying that
the single-lined fit presented above effectively removed the orbital signal so that the
spot-induced signal could be studied.  

The model we explored was composed of two stars whose orbital parameters are taken from
\citet{galli_gouldtextquotesingles_2018}.  For each star, we created a model spectral
line profile shifted in RV as dictated by the time of observation and orbital parameters.
The line profiles for each component were added together to create a final ``observed"
line profile which was analyzed in the same manner as the real observations to determine
the RV for the observation.  In order to add the profiles of the two stars we required the continuum flux ratio of the binary components, the intrinsic spectral line depth
of each star, and the $v$sin$i$ of each star.  We address each of these below and present
our final orbital modelling in which we fit the spot-subtracted RV variations
shown in Figure \ref{fig:rv_curve_model}.  While the $v$sin$i$ has been determined previously \citep[14.6 $\pm 1.6$ km s$^{-1}$][]{johns-krull_testing_2004},
we present a new determination of this
quantity below.  However, because we were not able to separate the spectral lines of the
two components, we were not able to determine individual $v$sin$i$ values for the two
stars.  We assumed that the one value we determine (likely an overestimate  given
the line blending) 
corresponds to the $v$sin$i$ of both stars.  We were able to self-consistently model the RV variations of Figure \ref{fig:rv_curve_model} using the known orbital
parameters from \citet{galli_gouldtextquotesingles_2018}, suggesting that our treatment
of $v$sin$i$ did not cause serious discrepancies. 

\subsection{HST Imaging of the Binary} \label{sec:imaging}

We relied on the HST imaging to determine the flux ratio of the two stars in the
Hubble 4 system.  In order to measure the relative flux ratios 
in the 11 HST WFC3 filters, we applied a least-squares fitting approach to 
the individual images. The parameters we varied were the horizontal and 
vertical positions of the components and their relative flux. Given the 
proximity of the components, there was a high degree of degeneracy between 
the relative stellar fluxes of the two and their positions. In order to resolve the 
degeneracy, we established a center of mass analog, which we call the center 
of flux. The center of flux in each of the filters was determined by a single 
source fit with an oversampled PSF including a Gassian blur in one of the two
shutter positions because of the short exposure times (see \S \ref{sec:data_img}).  
We varied the position and overall flux of a single source using our PSF, which
was then binned to the
observed pixel scale in order to create a model image.  We then determined 
the position which minimized the squared difference between the model and 
observed image.

\begin{figure}
    \centering
	\includegraphics[origin=c, angle = 90, width= 0.85\linewidth]{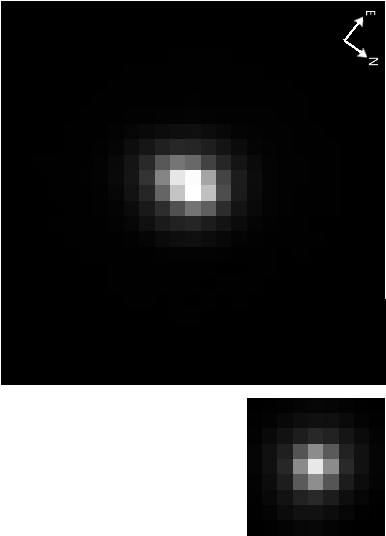}
	\caption{The third exposure taken on November 11, 2011, at UT 15:29:45 with the F775W filter on WFC3. The shutter is in the B position. The pixel scale is 0.04$^{\prime \prime}$ and the entire image is $25 \times 25$ pixels, or $1^{\prime\prime} \times 1^{\prime\prime}$. The position angle of the binary is 14$^\circ$. The Gaussian blurred PSF is shown in the upper right for reference, downsampled to match the resolution of the image.}
	\label{fig:h4_phot}
\end{figure}

\begin{deluxetable*}{ccccDD}
	\caption{Hubble 4 Photometric Model Results}	\label{tab:model}

	\tablewidth{0pt}
	\tablehead{
		\colhead{} & \colhead{Exposure Time} & \colhead{Separation} 
		& \colhead{Position Angle} & \colhead{} & \colhead{}\\
		\colhead{Filter} & \colhead{(s)} & \colhead{(mas)} & \colhead{(deg)} & \multicolumn2c{Flux Ratio} & \multicolumn2c{Final Flux Ratio}
	}
	\startdata
	F275W & 100 & $61^{+8}_{-12}$ & $5^{+15}_{-17}$ & $0.281^{+0.431}_{-0.258}$ && $0.323^{+0.117}_{-0.099}$ \\
	F336W & 25 & $65^{+6}_{-12}$ & $18^{+8}_{-7}$ &  $0.309^{+0.288}_{-0.199}$ && $0.351^{+0.080}_{-0.071}$ \\
	F390W & 8 & $65^{+5}_{-8}$ & $18^{+5}_{-6}$ &  $0.411^{+0.246}_{-0.183}$ && $0.495^{+0.081}_{-0.074}$ \\
	F395N & 100 & $61^{+3}_{-8}$ & $14^{+7}_{-6}$ &  $0.389^{+0.259}_{-0.185}$ && $0.393^{+0.094}_{-0.083}$ \\
	F438W & 4.5 & $70^{+5}_{-11}$ & $20^{+6}_{-7}$ &  $0.440^{+0.376}_{-0.248}$ && $0.556^{+0.098}_{-0.087}$ \\
	F475W & 1.0 & $64^{+3}_{-5}$ & $12^{+2}_{-3}$ &  $0.549^{+0.109}_{-0.076}$ && $0.587^{+0.028}_{-0.027}$ \\
	F555W & 0.48 & $71^{+3}_{-5}$ & $17^{+5}_{-5}$ &  $0.722^{+0.378}_{-0.269}$ && $0.801^{+0.082}_{-0.076}$ \\
	F625W & 0.48 & $71^{+4}_{-5}$ & $15^{+4}_{-3}$ &  $0.734^{+0.555}_{-0.331}$ && $0.829^{+0.080}_{-0.074}$ \\
	F656N & 9.0 & $71^{+6}_{-9}$ & $18^{+6}_{-7}$ & $0.539^{+0.604}_{-0.340}$ && $0.693^{+0.107}_{-0.095}$ \\
	F775W & 0.48 & $69^{+5}_{-6}$ & $14^{+4}_{-3}$ & $0.741^{+0.134}_{-0.179}$ && $0.817^{+0.052}_{-0.049}$ \\
	F850L & 0.80 & $65^{+3}_{-3}$ & $14^{+3}_{-2}$ &  $0.773^{+0.135}_{-0.083}$ && $0.781^{+0.037}_{-0.036}$ \\
	\enddata
\end{deluxetable*}

The two-source fitting via least squares with an oversampled PSF was then combined with 
the center of flux determined from the single source fit and the 
separation between the sources in the horizontal and vertical directions was varied. 
A region of $9 \times 9$ WFC3 pixels was cut out of each image around the center
of flux location.  The CCDs in the UVIS channel of WFC3 have pixels that
are 0.04$^{\prime\prime}$ on a side, so the extracted regions were 
0.36$^{\prime\prime}$ on a side.  The nominal FWHM of the UVIS PSF is
$\sim 0.075^{\prime\prime}$ or narrower in the bands observed.  We first
normalized the observed image subregion by the total flux because
the PSFs used were normalized to unit total flux.  Each of the two sources
was assigned a relative flux, the brighter given by $F$, 
$0.5\leq F \leq 1.0$, and the dimmer given by 1-$F$. Then shifting the 
brighter source by ((1-$F$)$dx$,(1-$F$)$dy$) relative to the center of flux and the 
dimmer source by (-$Fdx$,-$Fdy$) relative to the center of flux, with $F$ 
greater than 0.50 to prevent degenerate solutions, the binary was 
reproduced by finding the best fit parameters $F, dx$, and $dy$ which
minimized $\chi^2$ between the model and observed images. 
Table \ref{tab:model} gives the results of the least-squares 
fitting for the 11 filters, where the flux ratio is defined as $(1-F)/F$ and is
the flux of the secondary divided by the flux of the primary.
The position angles (PA) of the secondary 
relative to the primary describe the projected position of the binary in the plane of the 
sky measured in degrees East of North.
The PAs and separations are relatively consistent across filters. 
The mean separation and PA (and their mean uncertainties) of the components 
are found to be $66.7 \pm 7$ mas and $15^{\circ} \pm 6^{\circ} $ respectively. 
For the date of the HST observations, the orbital parameters from 
\citet{galli_gouldtextquotesingles_2018} predicted a separation of $62.86\pm 1.27$ mas
and a position angle of $13.98^{\circ} \pm  5.14^{\circ}$, showing that our results 
are fully consistent the \citet{galli_gouldtextquotesingles_2018}. 

\begin{figure}%
	\centering
	{\includegraphics[angle = 90, origin=c,width=0.425\linewidth]{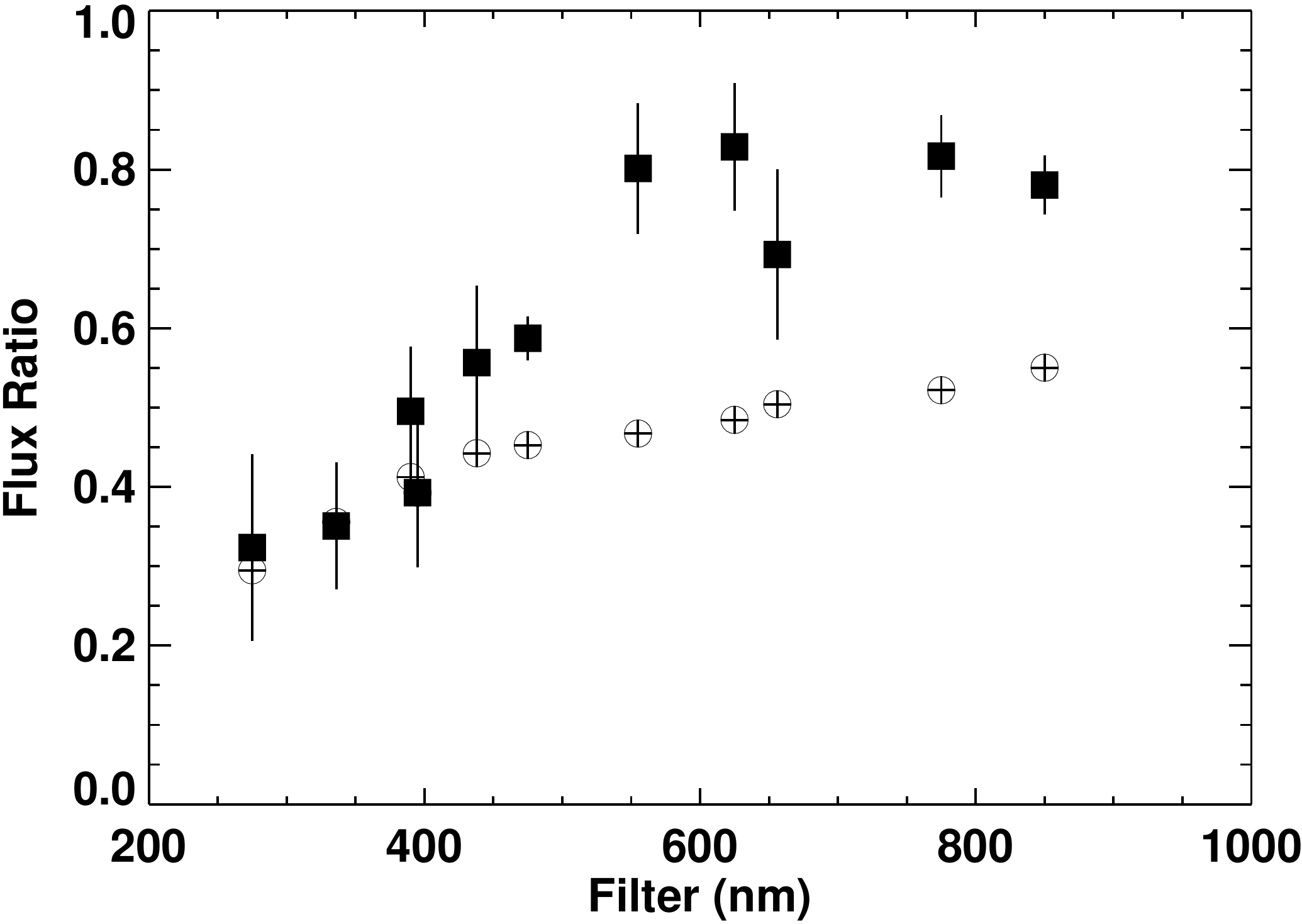} }%
	\qquad
	{\includegraphics[angle = 90, origin=c,width=0.425\linewidth]{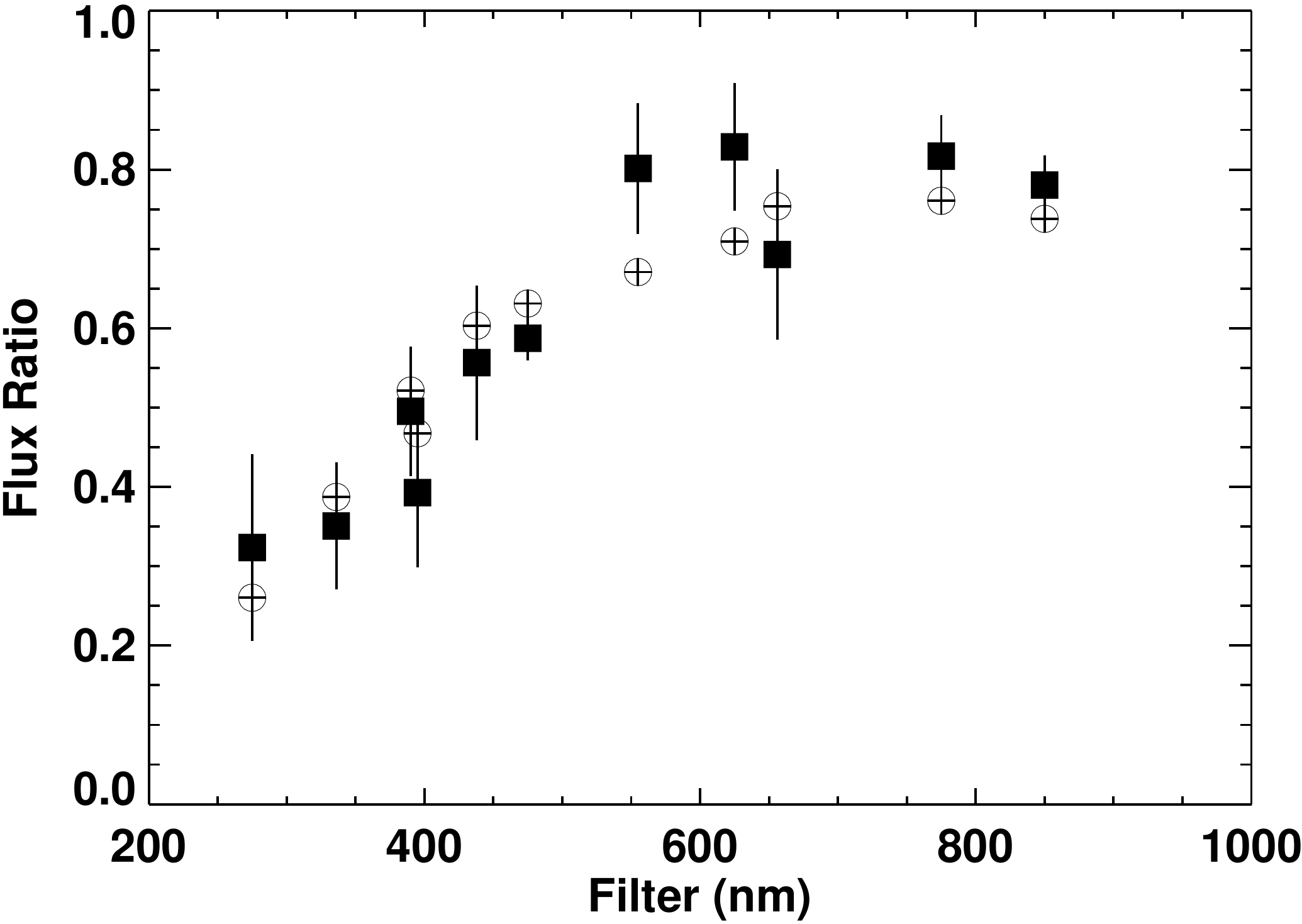} }%
	\caption{\textbf{Left:} Filled squares represent flux ratios from best fit of images, circles with crosses show flux ratios for NEXTGEN spectra assuming no spot coverage. \textbf{Right:} Filled squares represent flux ratios from best fit of images, circles with crosses show flux ratios for NEXTGEN spectra assuming 60$\%$ spot coverage of the primary, 0$\%$ spot coverage of the secondary. }%
	\label{fig:fluxratios}%
\end{figure}

In reality, the relative position of the two sources should not depend on
the wavelength of observation.  Since our separations and PAs are consistent 
with \citet{galli_gouldtextquotesingles_2018}, we analyzed the images again,
holding these parameters fixed at the \citet{galli_gouldtextquotesingles_2018} values, and solve only for the flux ratio.
The resulting ratios are given in the last column of Table \ref{tab:model} and are plotted
in Figure \ref{fig:fluxratios}.  Generally, the two stars are more similar
in brightness at redder wavelengths than they are in the blue, indicating
that the secondary component is cooler than the primary, as expected for the more massive
primary.  The flux 
ratios in the visible filters (F555W, F625W) have an interpolated value 
of $0.80 \pm 0.08$ at the mean wavelength used for the radial velocity 
analysis (5594 \AA) described below.

To test how these flux ratios compare with those predicted by current young
star models, we took stellar parameters from \citet{baraffe_new_2015} 
isochrones using the masses from \citet[][and reported in Table \ref{tab:h4parms}]{galli_gouldtextquotesingles_2018}
at an assumed age of 2 Myr.  These models give effective temperatures and 
radii of 4530 K and 2.1 R$_\odot$ for the primary
and 4070 K and 1.7 R$_\odot$ for the secondary.  We 
used synthetic spectra from the NEXTGEN model atmospheres 
\citep{hauschildt_nextgen_1999} to compute predicted flux ratios. 
Specifically, the BT-NextGen low resolution synthetic 
spectra\footnote{Obtained from  http://svo2.cab.inta-csic.es/theory/newov/} 
\citep[][]{allard_models_2012} with the 
\citet{asplund_chemical_2009} abundances at a variety of 
temperatures.  We varied the effective temperature between 3800K to 4300K for
the secondary and  between 4500K to 4800K for the primary to estimate the 
predicted flux ratios. The synthetic spectra were multiplied by the HST 
WFC3/UVIS passbands\footnote{Obtained from  http://www.stsci.edu/hst/instrumentation/wfc3/performance/throughputs} \citep{kalirai_wfc3_2009}, 
integrated, and scaled according to the assumed radii for each component.
The flux ratio of the system in each of the bands was then determined for 
the assumed temperatures.  While the extinction to Hubble 4 is generally
considered to be low, we assumed it was the same to each component and
thus did not take it into account to estimate the flux ratios.
The combination of temperatures which 
yielded the most similar flux ratio curve to that of the observations is 
4500K for the primary and 4300K for the secondary (left 
panel of Figure \ref{fig:fluxratios}).  Clearly, the model flux ratio curve is discrepant with respect to the observations, and
the parameters have each gone to limits we imposed to force the temperatures of the stars as close together as the fitting allowed.
This is because the ratio of surface areas for the stars is 0.66, which
is then the largest the flux ratio could be if the stars had the same
temperature.  As long as the less massive star is cooler, the ratio will
be smaller, and as a result will not provide a good fit to the observed
flux ratios.  Either the relative radii of the two stars are poorly 
predicted by the \citet{baraffe_new_2015} models, or some other effect
must be considered.

To attempt to better approximate the flux ratio curve, we allowed 
the two components to have large starspots.  We
used NEXTGEN spectra for the predicted effective temperatures, 4530K and 4070K \citep{baraffe_new_2015}. We assumed the spot temperatures are 1000K less 
than those of the photospheres \citep[e.g.,][]{bouvier_coyotes-i_1993} and added these 
to the spectra of the components at varying filling factors.  We held the 
unspotted photospheric temperatures at the values found above.  This results
in a much better fit, shown in the right panel of Figure \ref{fig:fluxratios}.
The resulting best fit spot coverage was 60$\%$ on the primary with the
secondary unspotted.  While this was crudely estimated, it appears that 
pre-main sequence models can reproduce the HST photometry if we allow for large spot coverage on at least the primary. Although it is unlikely that the secondary is entirely unspotted, the dominant flux of the primary drives the fit to suppress any spot contribution from the secondary. The assumption of large spot coverage on the primary explains the greater flux ratios found in the redder wavelengths here, as well as in the NIR as discussed in \citet{rizzuto_dynamical_2020}.  

\subsection{Optical vsini Measurement} \label{sec:vsini}

We attempted to disentangle the RV signals resulting from the binary motion from
those produced by starspot by modeling the observed
line profiles as the sum of spectral lines from the two stars.  Such
modeling requires an estimate of the $v$sin$i$ of the two sources.  As
the spectral lines from the two stars are far from resolved, we estimated
a single $v$sin$i$ value by analyzing the observed profiles as if they come from
only one source and then assumed that the two stars have the same
$v$sin$i$.  Our resulting estimate of the $v$sin$i$ is therefore likely an overestimate compared to the true rotation rate.  

The $v$sin$i$ values were measured using the same general technique outlined
in \citet{hartmann_rotational_1986} and \citet{soderblom_calibration_1989}. The premise of this method
is that the full width at half maximum (FWHM) of an appropriate mean
line profile is directly proportional to the $v$sin$i$ of the star.  In many
cases, the mean line profile used is the CCF, created when the target star spectrum
is cross-correlated against an unbroadened, narrow-lined spectrum.  
Another suitable mean line profile that can be used results from applying the
technique of Least Squares Decomposition \citep[LSD;][]{donati_differential_1997}
to dramatically boost the signal to noise of individual observations.
For this paper, LSD line profiles were constructed using the line list and code
described in \citet{chen_spectropolarimetry_2013}.  Using this method to
measure $v$sin$i$ requires that we first calibrate the relation between the 
FWHM of the LSD profile and the $v$sin$i$.  In order to do this, we used the
observed spectrum of a slowly rotating star of similar spectral type to
Hubble 4.

HD 65277 is classified as a K4 dwarf star \citep{houk_michigan_1999}. Using 
the spectral type to effective temperature calibration for dwarf stars from 
\citet{pecaut_intrinsic_2013}, this spectral type implies an effective 
temperature of 4620 K. \citet{valenti_vizier_2005} included HD 65277 in 
their spectroscopic analysis of 1040 F, G, and K dwarfs, and found an effective 
temperature of 4741 K and a $v$sin$i$ = 1.0 $\text{kms}^{-1}$. The recent 
spectroscopic analysis of \citet{soto_spectroscopic_2018} identified a stellar 
$T_{eff}$ = 4660 $\pm$ 14 K for HD 65277 along with a rotational velocity 
of $v$sin$i$ = 1.81 $\pm$ 0.08 $\text{km s}^{-1}$. As these 
estimates of the rotational velocity are below the resolution (2.5 
$\text{km s}^{-1}$) of our optical data, HD 65277 serves as essentially 
a non-rotating template with a spectral type very similar to that of Hubble 4.

To construct the relationship between the FWHM of the LSD profile and the 
stellar $v$sin$i$, we artificially broadened the observed spectrum of HD 
65277 with a standard rotational broadening kernel 
\citep{gray_observation_2005}. We then computed the LSD profile of this 
rotationally broadened spectrum and measured the FWHM of the LSD profile. 
This procedure was repeated for several values of $v$sin$i$ to produce a 
calibration curve for the data. We then used the Hubble 4 LSD profile of 
each observation and linearly interpolated on the calibration relationship
to get a $v$sin$i$ measurement for each observation.  We then took the
mean of the different $v$sin$i$ values as our final estimate of the 
rotational broadening. We also computed the standard deviation of the mean 
for the multiple $v$sin$i$ values and assumed this as the random uncertainty 
associated with our measurement.  We estimated the systematic uncertainty in 
our analysis by using two additional slowly rotating stars \citep[HD 80367 and
HD 88371 with $v$sin$i$ values of 1.9 and 1.4 km s$^{-1}$, respectively;][]{valenti_vizier_2005}
as templates. We achieved agreement between the $v\text{sin}i$ measurements made with the three templates to typically better than
$0.7$ $\text{kms}^{-1}$, which we took as our systematic uncertainty and 
added in quadrature to the random uncertainties.  Our final $v$sin$i$
estimate for Hubble 4 was $14.0 \pm 0.8$ km s$^{-1}$, which we assumed
to be equal for both the primary and the secondary. This value is within errors of other previous measurements \citep[e.g., ][]{johns-krull_testing_2004}. 

\subsection{Modeling of the Hubble 4 Orbital RV Variations} \label{appendix:binary_flux_fit}

We modeled the orbital contribution to the RV variations of 
Hubble 4 by constructing synthetic line profiles of the system and analyzing them 
in the same way as the observations.  We created a template line profile for
each component of the binary and added them together, taking into account 
continuum brightness differences and RV offsets of the two components to
create each epoch's ``observed" line profile.  The data we try to reproduce are 
the initial spot subtracted RV measurements shown in Figure \ref{fig:rv_curve_model}. As in the spot fits, we omit the 2017 data from those used to constrain the double-lined binary model. The LSD line profile of HD 65277 was used as 
the template to model the spectral line profiles of each component.

\begin{figure}
    \centering
	\includegraphics[angle = 90, origin=c, width= 0.85\linewidth]{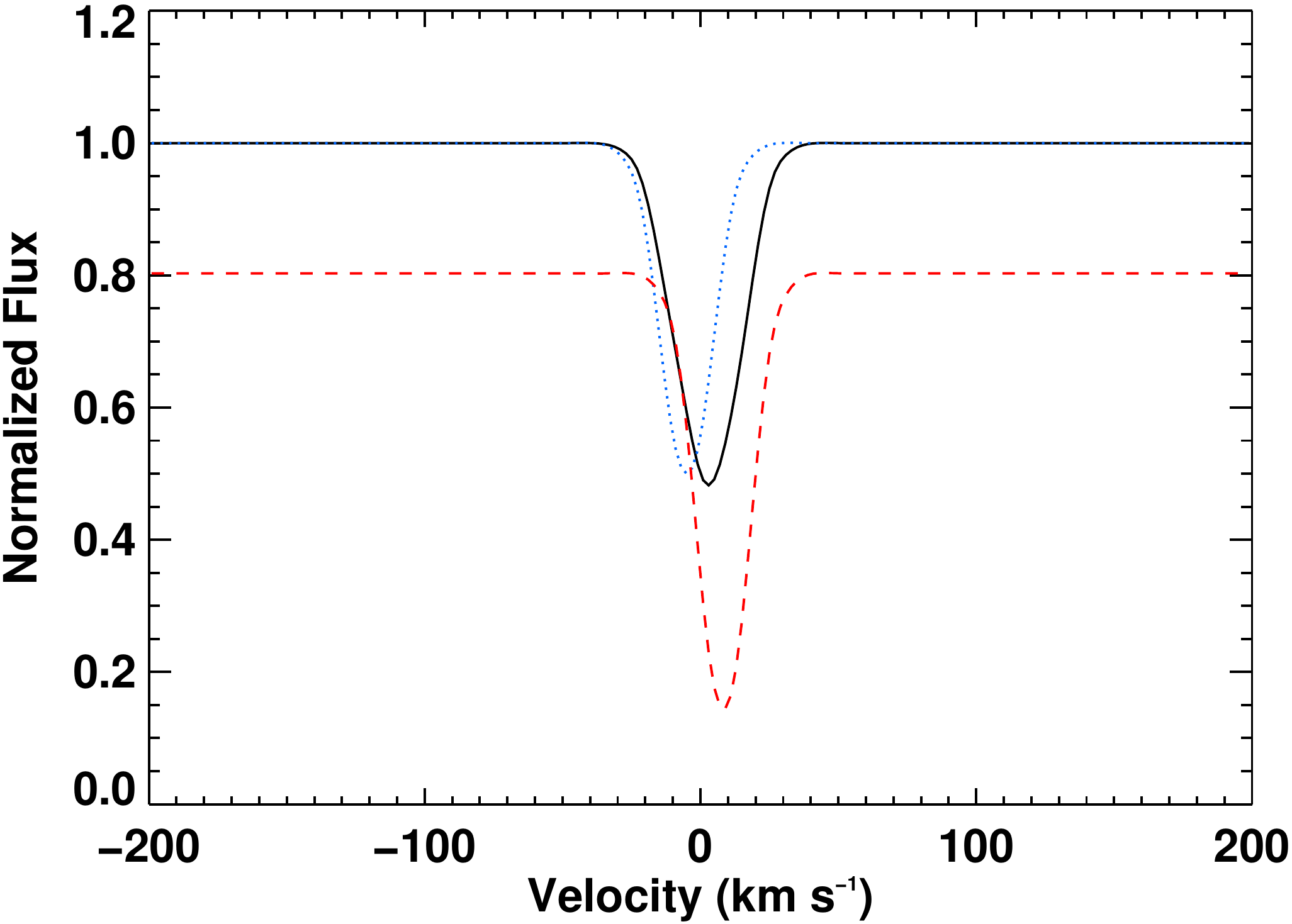}
	\caption{A sample model line profile generated using LSD profiles. The secondary star is shown with a dashed line (red), the primary is shown with a dotted line (blue), and the combined and renormalized model shown in solid black.}
	\label{fig:sample_model}
\end{figure}

In order to model the binary, the predicted RV values of each component, based
on the orbital parameters given in Table \ref{tab:h4parms}, were computed at each
observing epoch.  We assumed the two stars have the same $v$sin$i$, 14.0 km
s$^{-1}$, and then rotationally broadened the template by the this amount.
We made two copies of the broadened template, and shifted the broadened template
by the RV values of the two stars (Figure \ref{fig:sample_model}).
Leaving the Hubble 4 A profile normalized to 1.0 in the continuum, the Hubble 
4 B profile was then multiplied by a flux ratio, $f$. The two profiles 
were then added together and renormalized by $1+f$, and all of the resulting 
profiles were cross-correlated with the single rotationally broadened template
to measure the resulting RV value.  These model RVs were then fit to the
observed RVs, and reduced $\chi^2$ minimization was used to find the 
flux ratio which best fit the observations, resulting in 
$f = 1.07_{-0.04}^{+0.05}$. This differs significantly
from the best fitting flux ratio (0.80) in the HST photometric data 
discussed above and implies the secondary is {\it brighter} 
than the primary.  We then tried fixing the continuum flux ratio to 0.80 
and then varied the ratio of template line 
depths between the primary and secondary component and refit the observations.
We found a line depth ratio of $0.61 \pm 0.03$ 
(i.e. the lines of the primary are 61$\%$ as deep as those of the secondary)
brings the RV model in agreement with the imaging data. Deeper lines in the secondary better match expectations. The cooler chromosphere of the secondary would absorb more of the blackbody core emission than the hotter primary's chromosphere. An example set of
line profiles used in the final 
modeling is shown in Figure \ref{fig:sample_model}. The orbital
RV model fit based on the LSD analysis is shown in Figure \ref{fig:rv_curve_model}. 

To investigate whether the final line ratio we found for this system is 
reasonable, we used the Vienna Atomic Line Database 
\citep{piskunov_vald_1995, ryabchikova_major_2015} to extract spectral line 
data and synthesize models of the two components, assuming effective 
temperatures of 4400K for component A and 4000K for component B (these 
temperatures correspond to the models in the NextGen grid that are
closest in effective temperature to the predicted temperatures found
for each star).  A surface gravity (log~$g$) of 3.5 is assumed, as described in 
\citet{johns-krull_testing_2004}. We used the synthesized spectra to
compute LSD line profiles for the two effective temperatures and employed the VALD
line list covering the same wavelengths used to measure the RV and bisector
span for the actual observations.  The spectra synthesized from these models 
indicated a line depth ratio of 0.91.  Although this ratio is in the correct
sense (the lines of the primary are shallower than in the secondary), the
value does not exactly match that necessary to fit the orbital model to the RV
data.  However, the synthesized spectra are computed without knowing the exact 
temperatures and gravities of the component stars.  In addition, the final
RV model used assumes a component flux ratio determined from the HST
observation which is from a very different epoch compared to all the
RV measurements in Figure \ref{fig:rv_curve_model}.  As a result, this line depth ratio may not be exactly appropriate
considering that the degree of spottedness on the component stars changes as the result of
stellar rotation and likely has longer term changes with time.   

\subsection{Comparison with Previous Studies and Mass Limits on a Possible 
Third Component}

\citet{rizzuto_dynamical_2020} performed a similar analysis on the same set of HST
images of Hubble 4 that we presented in this paper. To account for the shutter-induced jitter in shorter exposures, they apply a Gaussian blur to the PSF used in both the
A and B shutter position, allowing both the extent and angle of the blur to vary in
their fits.  We apply a symmetric blur only in the B shutter position based on 
calibration data from \citet{sabbi_uvis_2009}.
The flux ratio measurements are sensitive to the positions of the components relative
to the size and orientation of the blur. Based on the size of the residuals (around
$\sim 5$\% of the peak flux for both studies), our fits are of similar quality to
those of \citet{rizzuto_dynamical_2020}. 

Possibly as a result of the difference in PSFs used, the flux ratios reported by \citet{rizzuto_dynamical_2020} differ significantly from those reported in Table \ref{tab:model}. The greatest difference is in F555W, where our measured flux ratio of $0.801^{+0.082}_{-0.076}$ differs significantly from their measurement of $0.402^{+0.083}_{-0.104}$. Though the F555W measurement was omitted from their later analysis, in the 5 filters redder than 500nm, the flux ratios reported here are consistently larger than those measured by \citet{rizzuto_dynamical_2020}. \citet{rizzuto_dynamical_2020} also report a greater-than-expected flux ratio of 0.65 measured in the NIR, as well as the presence of TiO lines in the unresolved spectrum of the binary. The flatter SED ratio reported in \citet{rizzuto_dynamical_2020} is best fit by spectra 
for the two components with similar temperatures, with the optimal fit finding $T_\text{eff} = 4411$K for the primary and $T_\text{eff} = 4254$K for the secondary. These temperatures are similar to the results of our unspotted model presented in Figure \ref{fig:fluxratios}. However, as reported before, these temperatures do not match those expected for this system according to young star models \citep{baraffe_new_2015}. Although the best-fit spectra that assume no spot coverage match the NIR flux ratio of approximately 0.65, they do not predict the strong TiO features found in the unresolved spectrum of the binary. 

To explain these observations, \citet{rizzuto_dynamical_2020} propose that the Hubble 4 system is in fact a hierarchical triple system. They suggest the ``primary" of the resolved 
binary is in fact itself a close binary system in which $T_\text{eff}$ is $\sim$ 4000K for the more massive component.   We use our RV measurements to place an upper limit on the mass of an unseen companion orbitting Hubble 4 A following the procedure described in \citet{boyajian_planet_2016}, which we summarize here.  In this search we use the
RV variations after subtracting the known binary contribution using the model described above in \ref{appendix:binary_flux_fit} and also the fitted spot models described in \S \ref{sec:spot rvs}. We phase-fold the residual RVs to $10^6$ periods, ranging from 2 days to 1000 days. We then fit a sinusoid to each resulting RV curve via least squares to determine the RV amplitude and uncertainty, $\sigma$. We use the measured RV amplitudes and add 2$\sigma$ to get a conservative upper limit for the maximum allowed RV amplitude at each period. 

\begin{figure}
    \centering
	\includegraphics[angle = 90, origin=c, width= 0.85\linewidth]{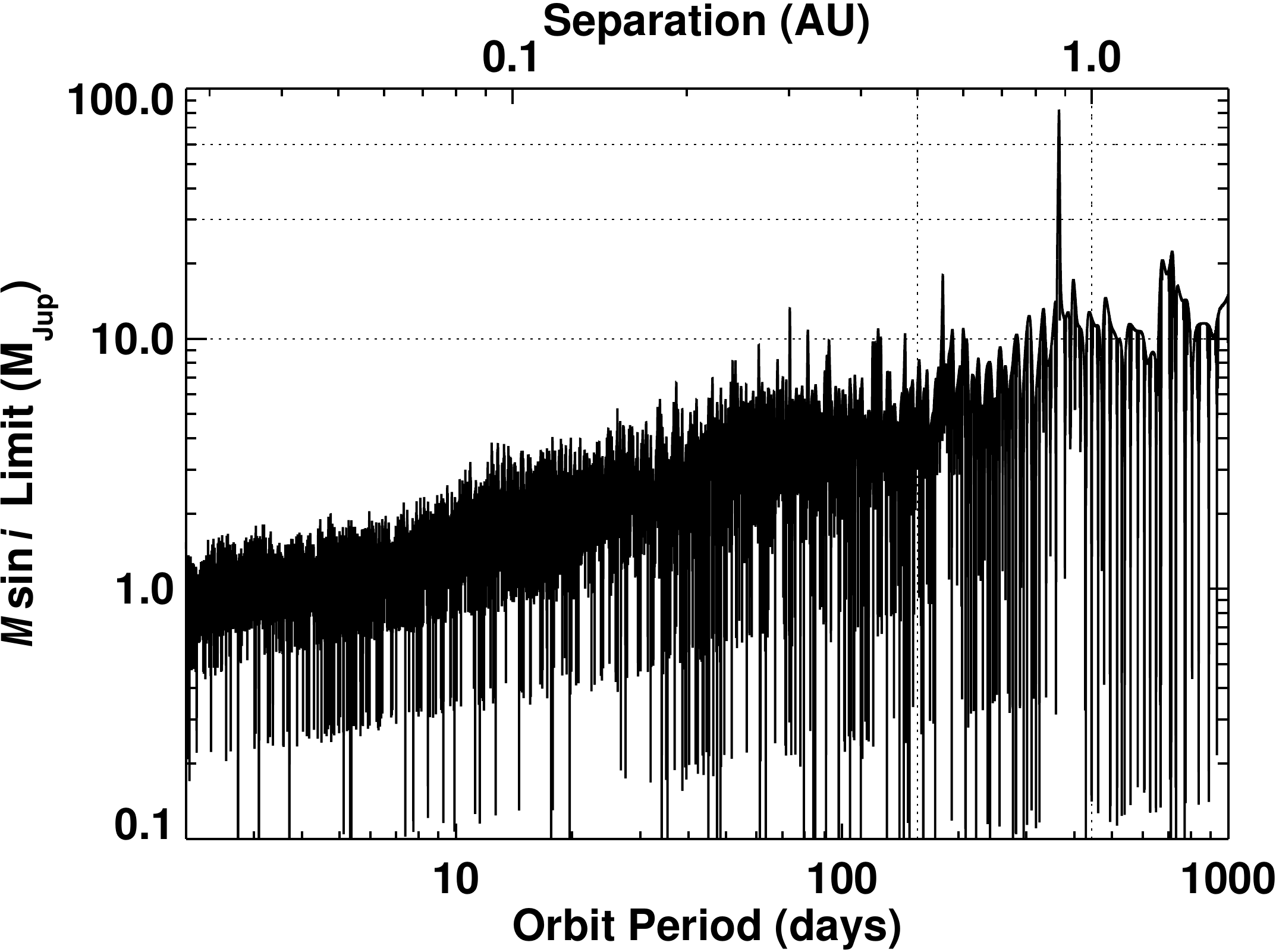}
	\caption{Mass limits on undetected companions around Hubble 4 A. Vertical dotted lines at 0.5 AU and 1 AU show orbital separations, and horizontal dotted lines indicate companion masses of 10 $M_\text{Jup}$, 30 $M_\text{Jup}$, and 60 $M_\text{Jup}$. The separations are computed for a test particle at the given orbit period.}
	\label{fig:h4_detlims}
\end{figure}

To convert the RV amplitude to a companion mass, we assumed $0.7 M_\odot$ for the mass
of the primary in Hubble 4 A, estimated by using the \citet{baraffe_new_2015} tracks
and again assuming a system age of 2 Myr for a star with $T_\text{eff}=4000$K. The mass limits are shown in Figure \ref{fig:h4_detlims}. \citet{rizzuto_dynamical_2020} gave a maximum projected separation of $\sim 0.5$ AU. Inside this limit, we find $M \sin i > 8 M_\text{Jup}$ exceeds our 2 $\sigma$ detection limit. \citet{galli_gouldtextquotesingles_2018} found the primary of the Hubble 4 system has
a mass of $\sim 1.2 M_\odot$.  If this is in fact two stars with the more massive having
a mass of $\sim 0.7 M_\odot$, the third star in the system must have a mass of $\sim 0.5
M_\odot$.  Combined with our 8 $M_{\rm JUP}$ companion mass limit, the orbit of such a
companion would have to have an inclination of $i \leq 0.92^\circ$. 

We also estimated the resulting SED ratios from the proposed triple system following the 
procedure described in \ref{sec:imaging} and compared the results to the HST photometry. 
We used NEXTGEN synthetic spectra and assumed the effective temperature of Hubble 4 B and
the more massive component of Hubble 4 A are both 4000 K, and we assumed 
$T_{\rm eff} = 3700$ K for the less massive component of Hubble 4 A.  We also assumed the 
radii of the primary components are $\sim 1.66 R_\odot$ and $\sim 1.48 R_\odot$, as given by \citet{baraffe_new_2015} isochrones, and we again assumed the radius of Hubble 4 B
is $1.7 R_\odot$. The resulting model SED is much flatter than the SED measured from the HST photometry and is a worse fit than either of the fits shown in Figure \ref{fig:fluxratios}. The predicted flux ratios are also much greater than those measured here or in \citet{rizzuto_dynamical_2020}, especially in wavelengths shorter than 500 nm. 

Our RV measurements, combined with the \citet{galli_gouldtextquotesingles_2018} orbit
determination of the wide binary, place tight restrictions on the parameter space 
allowed for a third member of the Hubble 4 system. We also find that the optical SED ratios
of the system measured from our analysis of the HST imaging of Hubble 4 does not
match well the predicted flux ratios for the hierarchical triple system proposed by
\citet{rizzuto_dynamical_2020}.  While our analysis of the HST data differs from
that of \citet{rizzuto_dynamical_2020}, their analysis of the HST data produces
flux ratios inconsistent the hierarchical triple system they propose, particularly at
shorter wavelengths.  Our measured HST flux ratios, along with the NIR flux ratios given
in \citet{rizzuto_dynamical_2020}, are more accurately reproduced by a model with large spot coverage on the $T_{\rm eff} = 4530$ K primary and an unspotted $T_{\rm eff} = 4070$
K secondary. The large spotted region on the primary could also be the source of the 
relatively strong TiO features observed in the spectra of the binary. Therefore, we 
concluded that a model with highly spotted primary and a relatively unspotted secondary,
with no significantly contributing third companion, provides a more plausible explanation for the Hubble 4 system.

\end{document}